\documentclass[12pt,onecolumn,twoside]{IEEEtran}
\usepackage{amsmath,amssymb,epsfig}
\usepackage{graphicx}
\usepackage{subfigure}
\usepackage[sort&compress,square,numbers]{natbib}
\newtheorem{theorem}{Theorem}
\newtheorem{proposition}[theorem]{Proposition}
\newtheorem{lemma}[theorem]{Lemma}

\newtheorem{assumption}{Assumption}
\newtheorem{claim}[theorem]{Claim}

\newtheorem{definition}{Definition}

\newcommand{\asnr}{\mbox{${\overline{SNR}}$}}
\newcommand{\ainr}{\mbox{${\overline{INR}}$}}
\newcommand{\aisr}{\mbox{${\overline{ISR}}$}}

\newcommand{\beq}{\begin{equation}}
\newcommand{\eeq}{\end{equation}}
\newcommand{\barrayl}{\begin{array}{lll}}
\newcommand{\earray}{\end{array}}
\newcommand{\bds}{\begin {itemize}}
\newcommand{\eds}{\end {itemize}}
\newcommand{\bdf}{\begin{definition}}
\newcommand{\blm}{\begin{lemma}}
\newcommand{\edf}{\end{definition}}
\newcommand{\elm}{\end{lemma}}
\newcommand{\bthm}{\begin{theorem}}
\newcommand{\ethm}{\end{theorem}}
\newcommand{\bprp}{\begin{prop}}
\newcommand{\eprp}{\end{prop}}
\newcommand{\bcl}{\begin{claim}}
\newcommand{\ecl}{\end{claim}}
\newcommand{\bcr}{\begin{coro}}
\newcommand{\ecr}{\end{coro}}
\newcommand{\bquest}{\begin{question}}
\newcommand{\equest}{\end{question}}

\newcommand{\pvec}{{\bf{p}}}

\newcommand{\Smat}{{\bf{S}}}

\newcommand{\E}{{\rm{E}}}

\newcommand{\be}{\begin{equation}}
\newcommand{\ee}{\end{equation}}
\newcommand{\beqna}{\setlength{\arraycolsep}{0.0em}\begin{eqnarray}}
\newcommand{\eeqna}{\end{eqnarray}\setlength{\arraycolsep}{5pt}}

\title{Competitive Spectrum Management with
Incomplete Information}
\author{Yair Noam, Amir Leshem, ~\IEEEmembership{Senior Member,~IEEE,} and  Hagit Messer,~\IEEEmembership{Fellow,~IEEE,}\thanks{Part of this paper will appear  in ICASSP 2010.}}
\usepackage{varioref}

\begin{document}

\maketitle

\begin{abstract}
An important issue in wireless communication is the  interaction  between selfish and independent wireless communication systems in the  same frequency band.  Due to the selfish nature of each system, this   interaction is well modeled    as  a strategic game where each player (system) behaves to maximize its own utility. This paper studies  an interference  interaction (game) where  each system    (player) has \emph{incomplete information} about the other player's  channel conditions. Using partial information,  players choose between frequency division multiplexing (FDM) and full spread (FS) of their transmitted power. An important notion in game theory is the Nash equilibrium (NE) which represents a steady point in the game; that is,  each player can only lose by unilaterally deviating from it. A trivial  Nash equilibrium point in this game is where players  mutually choose FS and interfere with each other. This point
may lead to poor spectrum utilization   from  a global network point of view and even for each user individually.

In this paper,
we provide a closed form expression for  a non  pure-FS $\epsilon$-Nash equilibrium point; i.e., an equilibrium point where players choose FDM for some channel realizations and FS for the  others.  To reach this point, the only instantaneous channel state information  (CSI) required by each user is its own interference-to-signal ratio.
We show that operating in this non pure-FS $\epsilon$-Nash equilibrium point increases each user's throughput and therefore improves the spectrum utilization, and  demonstrate that this performance gain can be substantial. Finally, important    insights are provided  into the behaviour of   selfish  and rational   wireless users as a function of the channel parameters such as fading probabilities, the interference-to-signal ratio.  \end{abstract}


\section{introduction}

Wireless communication has become increasingly popular in recent years since more and more communication systems share  the same band. Consider for example an urban area with wireless local access networks (LAN), bluetooth systems, cordless phone, etc. These systems create  interference  which results in   major performance loss. This is why,
 interference mitigation is such an important issue \citep[e.g.][]{Tse_paper,A_goldshmith_archive,leshem2007,Luo_and_hayashi,ji2007cognitive, yu2002,scutari2006,leshem2009game,SPM_Sep_2009}.

In wireless networks, interference can be high and the channel is time varying \citep[see e.g.][]{goldsmith2005wc}. Furthermore,  users may be  independent of each other and selfish in the sense that each one is only interested in maximizing its own utility. Thus, non cooperative game theory is an appropriate tool to analyze such interactions. An important notion in game theory is the  Nash equilibrium (NE) which represents a steady point the game; that is, the NE point is a strategy profile  which is the best response of each player given that the others do not deviate from it. As such, it can be self imposed on  network users who are selfish in nature.


\begin{figure}
\centering
\subfigure[Wireless interference channel.]{\label{figure_setup1}\psfig{figure=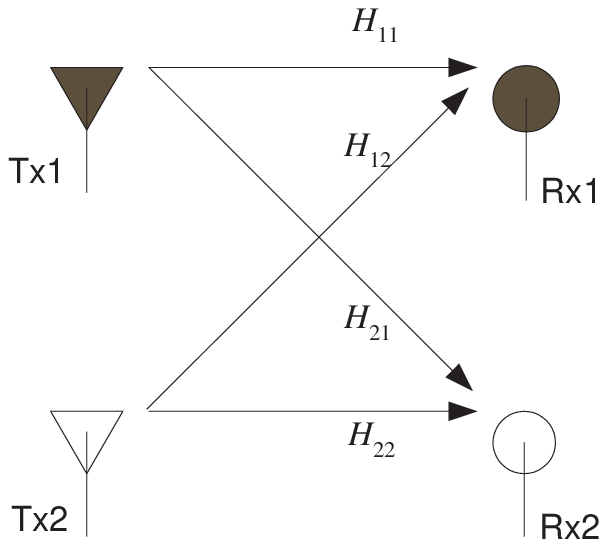,width=6.5cm}}
\qquad
\subfigure[Action space]{\label{figure_setup}\psfig{figure=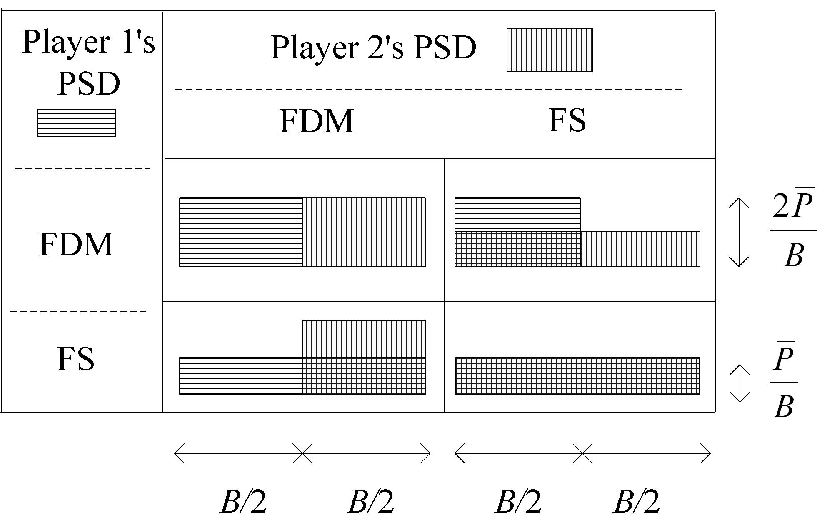}}
\caption{(a)  A wireless interference scenario with incomplete information. Each player knows the square magnitudes of its direct and impinging channel gains and the statistics of its  opponent's channel gains. For example player 1 knows $\vert H_{11}\vert^2$ and $\vert H_{12}\vert^2$ but knows only the statistics of $\vert H_{22}\vert^2$ and $\vert H_{21}\vert^2$. (b) the possible PSD configurations.  }
\label{figure_setup}
\end{figure}


 For  the case of a flat fading interference channel with full information\footnote{By complete information, we mean that every user knows all the direct and  cross channel gains of all users in the network.
} , it was shown \citep{Tse_paper} that Full Spread (FS) is a NE point, and a sufficient condition for its uniqueness was derived. It was further  observed that in many cases the FS NE point leads to  inefficient solutions. This happens    when mutual  FDM is better for both users than mutual FS but the system operates in a  mutual FS since the users are  subject to the prisoner's dilemma \citep{laufer_archive}.

The full information assumption is not always practical because  communicating channel gains between different users in a time varying channel within the channel coherence time  may lead to large overhead. In this case, it is more appropriate  to consider each channel coherence time as a one-stage game where players are only  aware of their own channel gains and their opponent's channel  statistics (which vary slowly compared  to the channel gains and therefore can be communicated \citep{leshem2007}). The interaction between the players may be repeated but with a different and independent channel realization each time and therefore is not a repeated game.
This motivates the use of games with incomplete information, also known as Bayesian games \citep{harsanyi1967gii,fudenberg} which have been   incorporated into wireless communications for  problems such as  power control    \citep{ heikkinen1999minimax,he2009bayesian,Jean2004Bayesian} and  spectrum management in the interference channel \citep{A_goldshmith_archive,NoamICASSP2010Bayesian_games}.  In \citep{he2009bayesian}, a distributed uplink  power control in   a multiple access (MAC) fading channel was studied and  shown to have  a unique NE point. This result however does not apply to the interference channel which is radically  different. In a MAC channel, user $i$'s  direct channel  gain is equal to the gain of the interference he creates  for the other users ($j\neq i$) while in the interference channel these parameters are independent. Thus, the interference channel is composed of a double number of parameters and   therefore is more complicated.

In this paper we analyze a two-user interference channel  with incomplete information in which each user
 knows the  magnitudes of its direct channel and of the impinging channel gains
 and its  noise power spectrum density (PSD) but is unaware of his opponent's  direct and impinging channel gains  but only knows it statistics (see Fig. \ref{figure_setup1}). Based on their measurements, users choose between pre-assigned FDM or FS (see Fig. \ref{figure_setup2}).
  This interaction may be repeated but a with different channel realization each time.

 With the same incomplete information, it was  shown \citep{A_goldshmith_archive}   that in a symmetric\footnote{Symmetric in the sense that the all channel gains (i.e. $H_{i,q}$ for every $i,q$) are identically distributed.}
  interference channel with a one-time interaction,  FS is the only  symmetric strategy profile\footnote{In symmetric strategy profile   users are restricted to identical strategies.} which is a NE point. This result however is limited to  scenarios  where all users  statistically   experience  identical channel conditions (due to the symmetry assumption) and does not apply to interactions between weak and strong users.

  A situation   where both players use  FS may lead to undesirable outcomes  from a global network point of view and even for each user individually.
 Thus, it is desirable to derive non FS Nash  equilibrium points which Pareto (which is component-wise larger) dominate the FS equilibrium point and lead to improved spectrum utilization.
A first step  toward this goal was made in \cite{NoamICASSP2010Bayesian_games} where it was shown that if users can coordinate in advance to use orthogonal FDM,  there exists a non pure-FS NE point which Pareto dominates the pure-FS NE point. This result however is also limited to symmetric interference channels.  This paper is aimed to fill this gap and  derive NE points in the general case of arbitrary channel distributions. For example, scenarios of weak and strong users (where one experiences a  high level of interferences and the other experiences low interference), different fading effects and cases where one has a strong line of sight path and the other has no line of sight. The assumptions  of arbitrary channel distribution together with the incomplete information that each user possesses about the other users in his vicinity are most suitable to the reality of  selfish users operating independently in  unlicensed frequency bands. This paper provides a closed form expression for  non pure-FS $\epsilon$-NE  point that increases each user's throughput and therefore improves the spectrum utilization, and  demonstrates that this performance gain can be substantial. The derived equilibrium point     provides insights into the the behaviour of   selfish and rational wireless users. Furthermore, it does not
require a central authority that imposes compliance of the
protocol. Thus, it provides  guidelines for designing a protocol that  users will choose voluntarily to follow.

The paper is orginazed as follows. In Section \ref{problem_formulation_section} we define the Bayesian Interference Game (BIG). This is  a two user interference interaction with incomplete information where the channel's  direct and crosstalks  gains are arbitrarily distributed (but independent). In Section \ref{best_response_section}  we present the best response function which is a user's best action  for his opponent's given strategy. We then provide a simple expression for the best response that depends only on the interference-to-signal ratio. In Section \ref{NE_points_of_the_BIG} we show (Proposition \ref{theorem that coopetation dominates}) that non pure-FS NE points  provide improved performance (with respect to pure-FS NE) to each user individually and therefore a better spectrum utilization. We then derive a closed form expression for  non pure-FS $\epsilon$ NE points.  Theorem \ref{theorem 3}  provides a sufficient condition for the existence of such points.

In section \ref{examples} we analyse the BIG in common wireless fading models (i.e. Rieghley and Rician and Nakagami) and learn the behaviour of selfish and rational wireless users in  various  wireless environments.

\section{problem formulation}
\label{problem_formulation_section}

\subsection{Notation and Definitions}
\label{two-user-game}
Consider a flat-fading interference channel  with two players, where
during a channel coherent  time, player i's  signal is given by (see Fig. \ref{figure_setup1})
\beq
W_i(n) = H_{ii}V_i(n) +H_{ij}V_j(n) + N_i(n)
\eeq
where $i,j\in\{1,2\},\; i\neq j$, $V_i(n), V_j(n)$ are user $i$'s and $j$'s transmit signals respectively, $N_i(n)$ is a white Gaussian noise with variance $\sigma_N^2$ and $H_{iq}, i,q\in\{1,2\}$ are the channel fading coefficients which are random variables. Throughout this paper, the index $j$ is never  equal to $i$. Both players have a total power constraint $\bar p$. We denote user $i$'s signal to noise ratio (SNR) and interference to noise ratio (INR) by $X_i=\vert H_{ii}\vert^2\bar p/\sigma_N^2$ and $Y_i=\vert H_{ij}\vert^2\bar p/\sigma_N^2$ respectively and denote   $\overline{SNR}_i={\rm E}\left\{X_i \right\}$, $\overline{INR}_i={\rm E}\left\{Y_i \right\}$.  We further denote the interference to signal ratio (ISR) by $Z_i=Y_i/X_i$. The realizations (sample points) of $X_i,Y_i,Z_i$ are denoted by $x_i,y_i,z_i$ respectively. When we want to stress that $x_i,y_i,z_i$ are the observed values of the SNR, INR and ISR they will be denoted by $SNR_i$, $INR_i$ and $ISR_i$ respectively.

\begin{assumption}
\label{assumption 1}
The channel gains  $\vert H_{iq}\vert^2, \;i,q\in\{1,2\}$  are continuous random variables with finite non zero moments and the probability density functions (PDF) \(f_{\vert H_{iq}\vert^2}(h)\), $i,q\in\{1,2\}$ are finite   for every $h>0$.
\end{assumption}


\subsection{The Bayesian Interference Game (BIG)}
In the BIG, user $i$'s channel state information (CSI) at the transmitter  side  are the realized values of $X_i$ and $Y_i$. It does not observe $Y_j$ and $X_j$  but only knows their  distributions.
 The channel is divided into two equal sub-bands  and  player 1's and 2's  actions  are given by
 \beqna
 \label{strategy_structure}
 \begin{array}{c}
 \pvec_1(\theta_1)=\bar p[\theta_1,\; 1-\theta_1]^T \\ \pvec_2(\theta_2)=\bar p[1-\theta_2,\; \theta_2]^T
 \end{array}
 \eeqna
 respectively (see Fig. \ref{figure_setup2}), where $\theta_i\in \Theta_i=\{1,1/2\}$ and  $\bar p$ is the total power constraint. The actions $\theta=1$ and $\theta=1/2$ correspond to FDM and FS, respectively.  This  formalism implies that  players coordinate in advance to use disjoint subbands in the case of FDM. This coordination can be carried out by using Carrier Sense Multiple Access (CSMA) techniques (see e.g. \citep{goldsmith2005wc}) where each player randomly chooses a subband and performs a random power backoff in case of collision. This is done at the first interaction when users exchange  information (channel statistics).

 We assume that during a single coherence period,  players manage their spectrum only once, based on their knowledge. Therefore, if the interaction is repeated it will be with different and independent channel realizations. This represents a case where the channel vary fast or a case where  simplicity requirements enable   a single spectrum shaping every coherence period.
 Player $i$'s utility function   $u_i(\theta_i,\theta_j,SNR_i,INR_i)$ is given in Table \ref{Table1}.
\begin{table*}[!t]
\centering \caption{User $i$'s payoff $u_i(\theta_i,\theta_j,SNR_i,INR_i)$ }
\begin{tabular}{|c|c|c|}
\hline
&player $j$ chooses FDM& player $j$ chooses FS\\
&$(\theta_j=1)$&$\theta_j=1/2$\\
\hline $\begin{array}{c}
$player $i$ chooses FDM$ \\
(\theta_i=1)
\end{array}$ & $\frac{1}{2}\log_2\left(1+
{SNR_i}\right)$&$\begin{array}{l} \frac{1}{2}\log_2\left(1+\frac{SNR_i}{1
+INR_i/2}\right)\end{array}$\\
\hline $\begin{array}{c}
$player $i$ chooses FS$ \\
(\theta_i=1/2)\\
\end{array}$ & $ \begin{array}{l}  \frac{1}{2}\log_2\left(1+\frac{SNR_i}{2}
\right)+\frac{1}{2}\log_2\left(1+\frac{SNR_i/2}{1+INR_i}
\right) \end{array}$&$\log_2\left(1+\frac{SNR_i/2}{1
+INR_i/2}\right)$\\
\hline
\end{tabular}
\label{Table1}
\vspace*{4pt}
\end{table*}
We are now ready to define the Bayesian interference game.

\begin{definition}
\label{BIG}
The Bayesian interference game (BIG) is defined by the following:
 \begin{enumerate}
 \item   Set of players $\{1,2\}$.
 \item   Action sets $\Theta_i=\{1,1/2\}$, $i=1,2$. Let $\theta_i\in\Theta_i$ be  the action chosen by player $i$, then according to (\ref{strategy_structure}),  $\theta_i=1$ corresponds to  FDM and $\theta_i=1/2$ corresponds to FS.
     \item A set of positive and independent random variables $X_1,\;Y_1,\;X_2,\;Y_2$ whose distributions are common knowledge. Each  player $i$ observes the realized values of  $X_i,Y_i$ but does not observe $X_j,Y_j$.
 \item A  utility function $u_i\left(\theta_i,\theta_j,x_i,y_i\right)$ given in Table \ref{Table1}.
 \item A set of pure strategies ${\cal S}={\cal S}_1 \times {\cal S}_2$ where every $S_i\in {\cal S}_i$ is a function that maps values of $x_i,y_i$  to an action in $\Theta_i$, i.e. $S_i:{\cal X}_i\times {\cal Y}_i\longrightarrow \Theta_i$, where ${\cal X}_i={\rm Range}(X_i)$ and ${\cal Y}_i={\rm Range}(Y_i)$.
\end{enumerate}

Player $i$'s objective is to maximize his conditional expected payoff given his private information  $x_i,y_i$, i.e.:\beq
\label{expected payoff}
\pi_i(S_i,S_{j},x_i,y_i)\triangleq\E\left\{ u_i(S_i,S_{j},X_i,Y_i)|X_i=x_i,Y_i=y_i \right\},\;\forall x_i,y_i\in {\cal X}_i\times{\cal Y}_i
\eeq
\end{definition}

 \begin{definition}
 a NE point  of the  BIG is a strategy profile $\Smat=(S_i,S_j)$ such that for every strategy profile $\tilde \Smat=(\tilde S_i,\tilde S_j)$ and every $i\in\{1,2\}$
 \beq
 \pi_i(S_i,S_j,x_i,y_i)\geq\pi(\tilde S_i,S_{j},x_i,y_i) \; \forall \;\; x_i,y_i\in {\cal X}_i\times{\cal Y}_i
 \eeq
 \end{definition}

 Since the action space is binary, a strategy $S_i(x_i,y_i)$ in the BIG is equivalent to  a decision region $D_i\subseteq {\cal X}_i\times{\cal Y}_i$ such that $S_i(x_i,y_i)=1$ (i.e. FDM) if $x_i,y_i\in D_i$ and $S_i(x_i,y_i)=0.5$ if $x_i,y_i\in D_i^c$. 

Two comments are in order:
\begin{itemize}
\item
Only pure strategies are considered in the BIG; that is, player $i$'s action is completely determined by his observed signal $x_i,y_i$. We do not consider mixed strategies which map values of  the observed signal $x_i,y_i$  to a probability distribution  on $\Theta_i$ i.e., player $i$ chooses randomly between FDM and FS with probability $a_i(x_i,y_i)$ and $1-a_i(x_i,y_i)$ respectively. A well known theorem in game theory is the Purification Theorem \cite[Theorem 6.2]{fudenberg}. It asserts that under some regularity conditions (among others  that each player's utility function  $u_i(\theta_i,\theta_j,x_i,y_i)$ should not be a function of $x_j$ and $y_j$), every mixed strategy has a pure strategy equivalent. Thus, all  NE points can be reached using pure strategies. The conditions of the Purification Theorem are satisfied in the BIG.
\item In the case where player $j$ chooses FDM,  FS is not the best action for player $i$. His payoff can be increased by performing waterfilling which will result in a higher rate. Therefore, it makes sense to modify the   FS action with the  waterfilling action  as considered in  \citep{laufer_archive,Tse_paper,yu2002} for interactions  with complete information. There are, however, two important caveats. The first is that the waterfilling solution in the interference channel must be carried out iteratively, where at  every iteration players measure their interference and shape their spectrum accordingly. The process needs to be repeated within the channel coherence time until convergence\footnote{ See  \cite{yu2002,Chung2003,Luo2005,scutari2006,Shum2007,pang_information_theory,leshem2009game} for further reference to the convergence of the iterative waterfilling procedure.}.  This may lead to large overhead in time varying channels and therefore is impractical. Moreover, the iterative waterfilling procedure does not necessarily converge  \cite{leshem2009game}. The second caveat is the analysis of the resulting game in the framework of incomplete information. The  result is a game with incomplete information where in addition to not knowing their opponent's utility,  players do not  know their own utility function  since it depends on their opponent's CSI. The analysis of such games is more complex and presents a greater challenge. For example,  the Purification Theorem is not satisfied if players use iterative water filling.

\end{itemize}

\section{Best response and approximate best response}
\label{best_response_section}
An important notion in game theory is the best response function. The best response function of player $i$ maps each of   player $j$'s strategies  to an action for which player $i$'s payoff is maximized. This function is used to derive NE points and is also important for  understanding   the  players' preferences and the nature of the game.

In this section we present   an expression for  the best response function of the BIG. This expression, however, is too complex for deriving a closed form expression for NE points of the BIG. Worse, it  does not provide insights into the game. For these reasons we obtain a simple approximation for the best response function  which provides greater insights into the game and will enable us to obtain a closed form expression for near NE points of the BIG.

\subsection{Best Response Function}
We now derive player \textit{i}'s best response to $S_j$ - player $j$'s strategy.  Note that $u_i(1,1,x_i,y_i)>u_i(1,1/2,x_i,y_i)$ since the  $\log$ is a monotonic function, and furthermore, due to Jenssen's inequality, $u_i(1/2,1/2,x_i,y_i)>u_i(1,1/2,x_i,y_i)$. Thus,
the following situations are possible:
\begin{itemize}
\item $A_i$ is the case in which $u_i(1,1,x_i,y_i)\geq u_i(1/2,1,x_i,y_i)$ which is equivalent to  $INR_i>SNR_i/2$
\item $B_i$ is the case in which $u_i(1/2,1,x_i,y_i)\geq u_i(1,1,x_i,y_i)$ which is equivalent to $INR_i\leq SNR_i/2$
\end{itemize}
Recall that
 player $i$ is not aware of the state of his opponent ($A_j$ or $B_j$) but only of his   probabilities.

 If player $i$ experiences  situation $B_i$ (which is $ISR_i\leq1/2$), then FS is his  best response. This  is because FS is a strongly dominating action; that is, it produces a higher payoff to player $i$ given any action of his opponent.
 It remains to find player $i$'s best response for  situation $A_i$;  i.e. the case where $ISR_i>1/2$, that is, strong interference.
Let $P(S_j=1)$ (the probability that player $j$ chooses FDM), then  player $i$'s payoff is given by
\begin{eqnarray}
\label{FDM_payoff}
\pi_i({\rm FDM_i},S_j,x_i,y_i)=P(S_j=1)u_i(1,1,x_i,y_i)+ (1-P(S_j=1))u_i(1,1/2,x_i,y_i)\\
\label{FS_payoff}
\pi_i({\rm FS_i},S_j,x_i,y_i)=P(S_j=1)u_i(1/2,1,x_i,y_i)+ (1-P(S_j=1))u_i(1/2,1/2,x_i,y_i)
\end{eqnarray}
Observe that player $i$'s payoff depends on his opponent's strategy and channel distribution only via $P(S_j=1)$;  hence the payoff will be denoted by $\pi_i(S_i,a_j,x_i,y_i)$ where
 \beq
 a_j=P(S_j=1)
 \eeq
 It follows that, player $i$'s best response is invariant to strategies with equal probability for  choosing FDM\footnote{ From player $i$'s point of view, ${\cal S}_j$ can be divided to into  equivalent classes ${\cal S}_{a_j}=\{S_j:P(S_j=1)=a_j\}$ such that ${\cal S}_{j}=\bigcup_{0\leq a_j\leq1}{\cal S}_{a_j}$.} and is dependent  on $S_j$ only via $a_j$.
 \begin{definition}
 \label{define_best_response}
Let $S_j$ be player $j$'s strategy with $a_j=P(S_j=1)$.
Player $i$'s best response to $S_j$  is  defined by:
\begin{eqnarray}
\label{best_response}
\check{S}_i(x_i,y_i,a_j)\triangleq\left\{\begin{array}{lc} \theta_i=1, & {\rm if }\;\; e(x_i,y_i,a_j)>0  \;{\rm and} \; y_i/x_i>1/2\\ \theta_i=1/2, & {\rm otherwise}
\end{array}\right.
\end{eqnarray}
\end{definition}

where
\begin{eqnarray}
\label{D1_region}
\begin{array}{ll}
 e(x,y,a)&=\pi_i({\rm FDM_i},a_j,x_i,y_i)-\pi_i({\rm FS_i},a_j,x_i,y_i)=\frac12 a\log(1+x_{i})-\frac a 2\log\left(1+\frac{x_{i}}{2}\right)\\&-\frac12\log\left(1+\frac{x_{i}/2}{1+y_{i}}\right)   - (1-a)\log\left(1+\frac{x_{i}}{1+y_{i}}\right)
  + \frac12 (1 - a)\log\left(1+\frac{2x_{i}}{1+y_{i}}\right)
  \end{array}
\end{eqnarray}

Note that  finding a NE point is equivalent to calculating $\hat a_1$ and $\hat a_2$ which solves the equations
\beqna
\label{NE eqaution}
\begin{array}{l} a_1=P\left(\check{S}_1(X_1,Y_1, a_2)=1\right)  \\
 a_2=P\left(\check{S}_2(X_2,Y_2, a_1)=1\right)
 \end{array}
\eeqna
and  that $a_1=0,\;a_2=0$ (pure-FS by both users) is a NE point regardless of the channel distribution since FS is the best response of each player if his opponent uses FS. In this case each player's payoff is $u_i(1/2,1/2,x_i,y_i)$. The pure-FS NE point may be very poor for both users as will demonstrated below.
\subsection{Approximate Best Response}

In order to analyze the best response function it will be simplified  by an  approximate best response. This approximate best response  plays in important role in deriving equilibrium points and understanding each player's preferences. The following proposition is needed before presenting the approximate best response.
\begin{proposition}
\label{proposition for q} Let
\beqna
\label{17cc}
\begin{array}{ll}
  r(a,q)=\frac{\log_2(q)}{2}-\log_2(1+q) +\frac{1}{2}\log_2(2+q)\\~~~~~~~~- a\left( 1+\log_2(1+q)-\frac{1}{2} \log_2(2+q)-\frac{1}{2} \log_2(1+2 q) \right)
\end{array}
\eeqna
then, for every $0<a \leq 1$ the following equation
\beqna
\label{17b}
 r(a,q(a))=0
\eeqna
has a unique solution $q(a)>1/2$ and therefore it defines an implicit function $q:(0,1]\longrightarrow (0.5,\infty]$. Furthermore, $q(a)$ is continuous and monotonically  decreasing.
\end{proposition}
\IEEEproof see Appendix \ref{appendix_a}.

\begin{definition}[approximate best response]
\label{approximate best response}
Let $S_j$ be player $j$'s strategy with $a_j=P(S_j=1)$.
Player $i$'s  approximate best response to $S_j$  is  defined by:
\begin{eqnarray}
\label{approximat best_response equation}
\tilde{S}_i(x_i,y_i,a_j)=\left\{\begin{array}{lc} \theta_i=1, & {\rm if }\;\; ISR_i=y_i/x_i> q(a_j)  \\ \theta_i=1/2, & {\rm otherwise}
\end{array}\right.
\end{eqnarray}
i.e. the approximate best response compares the ISR to a threshold $q(a_j)$\footnote{For $a=0$, we define $q(0)=\lim_{a\rightarrow 0} q(a)=\infty$. Under this definition, player $i$'s best response to the case where his opponent always chooses FS is to  choose FS.}.
\end{definition}

The intuition  behind the approximation is now described. First  consider the case of $SNR_i>>1$ (recall that $x_i,y_i$ are used interchangeably with $SNR_i,INR_i$ respectively). In this case
\beqna
e(SNR_i,INR_i,a)\approx \hat e(SNR_i,INR_i,a)
\eeqna
where
\begin{eqnarray}
\label{16}\begin{array}{lll}  \hat e(SNR_i,INR_i,a)\triangleq
 {\frac12 a_j\log(SNR_i) + \frac12 (1 - a_j)\log\left(1+\frac{2SNR_i}{INR_i}\right)} \\~~~~- (1-a_j)\log\left(1+\frac{SNR_i}{INR_i}\right)-\frac {a_j}  2\log\left(\frac{SNR_i}{2}\right)
 -\frac12\log\left(1+\frac{SNR_i}{2INR_i}\right)
 \end{array}
\end{eqnarray}
Thus,    $\check{S}_i(x_i,y_i,a_j)$  can be approximated by replacing $e(SNR_i,INR_i,a)$ with $\hat e(SNR_i,INR_i,a)$. Furthermore, note that
\begin{eqnarray}
\label{17}
  \hat e(SNR_i,qSNR_i ,a)=r(a_j,q)
\end{eqnarray}
 and recall that the equation $r(a,q)=0$ (see (\ref{17b})) defines the function $q(a)$.  Therefore,  $q(a)$ represents an ISR level for which FDM and FS yield approximately equal payoffs. Thus, if $SNR_i>>1$,    $\check{S}_i(x_i,y_i,a_j)$  can be approximated by a simple strategy which only compares the ISR to a threshold and choose action accordingly, i.e. it chooses FDM if
\beqna
\label{approximate_optimal strategy}
INR_i/SNR_i=ISR_i>q(a_j)
\eeqna
and chooses FS otherwise.

\begin{figure}
\centering
\psfig{file=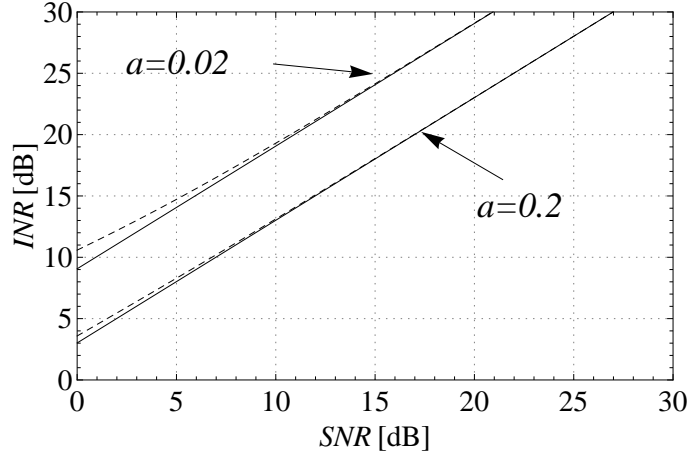,
width=0.5\textwidth} \caption{ Numerical evaluation of the best response function regions  given in (\ref{best_response}) for different values of $a_j$ ($a$ in the plot). For a given $a$,  points above the corresponding line belong to the FDM region.}
\label{Figure1}
\end{figure}

It remains to approximate  (\ref{best_response}) for the case where $SNR_i>>1$ is not satisfied.
 If $INR_i>>1$  and $ISR_i>1/2$ it can be shown that (\ref{best_response}) chooses FDM for every $0<a_j\leq 1$ and if $a_j=0$, it chooses FS. Thus, (\ref{approximate_optimal strategy}) is the best response  in this case as well since $ISR_i$ is greater  than $q(a_j)$ (which is finite  for every $0<a_j\leq 1$ and is infinite for $a_j=0$\footnote{under the convention that $\infty>\infty$ is  fuels. }).
 In the case of  $ISR_i\leq 1/2$, the best response in (\ref{best_response}) (which always chooses  FS because it is a strictly dominant strategy for player $i$) and the approximate best response  (\ref{approximate_optimal strategy}) coincide. This is because $q(a_j)\geq1/2$ for every $0\leq a_j\leq 1$.
 For the case where $INR_i>1/2 SNR_i$ but $INR_i$ and  $SNR_i$ are in the same magnitude as 1, the best response in (\ref{best_response}) cannot be simplified. However, numerical evaluation indicates that (\ref{best_response}) is well approximated by (\ref{approximate_optimal strategy}) as is depicted in Figure \ref{Figure1}.

We now present this idea formally.
To establish the  relation between the approximate and the  ordinary best responses,
define:  \beqna
 \label{observe check d}
 \check{D}^{a_j}_i&=&\{(x,y):e(x,y,a)>0, \;\; {\rm and}\; y>0.5x \} \\
 \tilde D^{a_j}_i&=&\{(x,y):y>q(a)x\}=\{(x,y):\hat e(x,y,a)>0,\;\; {\rm and}\; y>0.5x \}  \label{approximate D}
 \eeqna
  where  (\ref{approximate D}) is obtained by  substituting   $y=q x$ in (\ref{16}) and then invoking Proposition \ref{proposition for q}.
The following lemma describes precisely the sense in which $\tilde{S}_i(x_i,y_i,a_j)$ is approximately the best response. It shows that  in the high transmit power regime, the best response converges in probability to the approximate best response. Thus,  each  player is ``approximately"  indifferent to whether his opponent uses the approximate best response  or the true best response.

\begin{lemma}
\label{equivalence_condition}
Assume the channel gains  $\vert H_{iq}\vert^2, \;i,q\in\{1,2\}$  are continuous random variables,    then
for every $\epsilon>0$, there exist some $\overline{SNR}_0$  such that for every $\overline{SNR}_i>\overline{SNR}_0,\;i=1,2$  (or equivalently, for every $\bar p>\bar p_0$)\beqna
\label{decreases_to_zero}
P\left( \check{S_i}\left( X_i,Y_i,a_j\right) \neq \tilde{S_i}\left( X_i,Y_i,a_j\right) \right)<\epsilon
\eeqna
 furthermore, if   $\vert H_{iq}\vert^2, \;i,q\in\{1,2\}$  satisfy the regularity conditions in Assumption \ref{assumption 1},   $\epsilon$ decreases like
 \beqna\label{decreases like p square}
  \epsilon\leq O\left( \frac{\sigma_N^2}{\bar p^{2}}+\sum_{q=1}^2 F_{\vert H_{ii}\vert ^2}\left( \frac {\sigma^2_N}{\bar p^{1-\nu} }\right)F_{\vert H_{iq}\vert ^2}\left( \frac {\sigma^2_N}{\bar p^{1-\nu} }\right)\right)
 \eeqna
 for every \(0<\nu<1\).
\end{lemma}
\IEEEproof see Appendix \ref{equivalence_condition proof}.

\section{NE and $\epsilon$-NE Points of the BIG}
\label{NE_points_of_the_BIG}
 A trivial  NE point in the BIG is the pure-FS strategy profile . We would like to derive additional NE points which are non-FS.
   These points are of  interest because (if they exist) they  Pareto dominate  pure-FS NE points as  shown in the following proposition.
\begin{proposition}
\label{theorem that coopetation dominates}
Let $S_1, S_2$ be a non pure-FS NE point (i.e. $P(S_1=1),P(S_2=1) \neq 0$), then it Pareto dominates the pure-FS NE point, i.e. $\pi_i(S_i,S_j,x_i,y_i)\geq u_i(1/2,1/2,x_i,y_i)$ for all $x_i,y_i$ and $i$.
\end{proposition}
\IEEEproof See Appendix \ref{appendix_0}.

In the sequel, it is  shown that if users are allowed to coordinate in advance to use disjoint  subbands in the case of FDM (as implied in (\ref{strategy_structure})), FDM is possible from a game theoretic point of view and also increases the total system throughput as well as the individual throughput.

\subsection{Derivation of non pure-FS NE points}
\label{two user the general case section}

Proposition \ref{theorem that coopetation dominates} shows that non pure-FS NE points are attractive. However, deriving such points analytically is not always possible. For a symmetric game where all channel  magnitudes are identically distributed, NE points were derived in \citep{NoamICASSP2010Bayesian_games} where it was shown that in addition to the pure-FS NE point, there exists a non pure-FS NE given by the following strategy profile:
\begin{eqnarray}
\label{Nash_equilibrium} {S}_i(x_i,y_i)=
\left\{\begin{array}{lc} \theta_i=1, &  {\rm if }\;\;
y_i>x_i \\ \theta_i=1/2, & {\rm otherwise}
\end{array}\right.
\end{eqnarray}
However, in the general case of arbitrary distributions, NE points
 cannot be computed analytically.  This  makes them impossible to implement and analyze.  We therefore  address to near NE points.  

\begin{definition}
\label{define near optimal nash points} For $\epsilon\geq 0$,
an $\epsilon$-near  NE point is a strategy profile $(\hat S_1,\hat S_2)$ such that
\beqna\label{defination near optimal nash points equation 1}
 \pi_i\left(\hat S_i,\hat S_{j},x_i,y_i \right) \geq
\sup_{S_i\in {\cal S}_i} \pi_i\left( S_i,\hat S_{j},x_i,y_i \right)- \epsilon, \;\forall x_i,y_i
\eeqna
\end{definition}
It is straightforward to show that for sufficiently small $\epsilon$,  $\epsilon$-near NE points also Pareto dominate the pure FS NE point (this follows from the continuity of the expected payoff with respect to $a$).

The main idea behind $\epsilon$-near NE points is that if one of the players  deviates from it,  he can gain no more than $\epsilon$ additional payoff. From a practical point of view, for sufficiently small $\epsilon$,  $\epsilon$-near NE points are as stable as ordinary NE points.

We are now ready to introduce the main theorem which provides an analytic expression for such points:
\begin{theorem}
\label{approximate nash points representation theorem} Assume the channel gains  $\vert H_{iq}\vert^2, \;i,q\in\{1,2\}$  are continuous random variables,    then
for every $\epsilon>0$, there exists some $\overline{SNR}_0$  such that for every $\overline{SNR}_i>\overline{SNR}_0,\;i=1,2$  (or equivalently, for every $\bar p>\bar p_0$) the following strategy profile
is an $\epsilon$-near NE point: \beqna
\label{approximate nash points representation theorem - equation1}
\hat S_1=\check{S}_1(x_1,y_1,\hat a_2)\\
\label{approximate nash points representation theorem - equation2}
\hat S_2=\check{S}_2(x_2,y_2,\hat a_1)
\eeqna
 where   $\check{S}_i$ is the best response given in (\ref{best_response}),  and $(\hat a_1, \hat a_2)$ is a solution to the following equation system
\beqna
\label{a_1 curve}
a_1=1-F_{Z_1}\left(q(a_2)\right)
\\ \label{a_2 curve} a_2=1-F_{Z_2}\left(q(a_1)\right)
\eeqna
where $F_{Z_i}(z)$ is the distribution function of the ISR.
 Furthermore, if the channel gains  $\vert H_{iq}\vert^2, \;i,q\in\{1,2\}$  satisfy the regularity conditions in Assumption \ref{assumption 1} $\epsilon$ decreases like

 \beqna\label{decreases like p square}
  \epsilon\leq O\left( \frac{\sigma_N^2}{\bar p^{2}}+\sum_{q=1}^2 F_{\vert H_{ii}\vert ^2}\left( \frac {\sigma^2_N}{\bar p^{1-\nu} }\right)F_{\vert H_{iq}\vert ^2}\left( \frac {\sigma^2_N}{\bar p^{1-\nu} }\right)\right)
 \eeqna
 for every \(0<\nu<1\).

\end{theorem}

\IEEEproof see Appendix  \ref{approximate nash points representation theorem - proof}.

Theorem \ref{approximate nash points representation theorem} provides a procedure to calculate $\epsilon$-near NE points in the high averaged received SNR  regime. First, $\hat a_1$ and $\hat a_2$ are obtained by solving equations (\ref{a_1 curve}) and (\ref{a_2 curve}), then $\epsilon$-near NE points are given by (\ref{approximate nash points representation theorem - equation1}) and (\ref{approximate nash points representation theorem - equation2}).
 Each $\hat a_i$   is associated with a unique   threshold  \(\widehat{ISR }_i=q(\hat a_i) \) where above it FDM is approximately the best  strategy and below it, FS is the approximately the best strategy.

Although Theorem \ref{approximate nash points representation theorem} is proven rigourously in Section  \ref{approximate nash points representation theorem - proof}, we now explain it intuitively.
The idea behind the proof is to approximate player $i$'s best response $\check{S}_i(x_i,y_i,\hat a_j)$ by the simple approximate best response $\tilde S_i(x_i,y_i,\hat a_j)$ that satisfies  \beq\label{p8_22} P\left(\tilde{S}_i(X_i,Y_i,\hat a_j)=1\right) \approx P\left(\check{S}_i(X_i,Y_i,\hat a_j)=1\right)\eeq Note that the LHS  of \eqref{p8_22} can be expressed in closed form. This way, the equations in (\ref{NE eqaution}) are approximated by   (\ref{a_1 curve})-(\ref{a_2 curve}).
This enables us to obtain  $\hat a_1, \;\hat a_2$ analytically with the corresponding $\epsilon$-near NE point  given in (\ref{approximate nash points representation theorem - equation1}-\ref{approximate nash points representation theorem - equation2}).

\subsection{Existence of $\epsilon$-near NE Points}
Now that a procedure to derive $\epsilon$-near NE points has been established, we investigate the existence   properties of such points.
The following theorem presents a sufficient condition for the existence of a $\epsilon$-near  NE point.
\begin{theorem}
\label{theorem 3}
Assume that $Z_i$, $i=1,2$ are continuous random variables  such that $P(Z_i<0.5)<1$ and denote the corresponding densities by $f_{Z_i}(z)$. A sufficient condition for the  existence of a solution to equations (\ref{a_1 curve}), (\ref{a_2 curve}) is that
\beqna
\label{25}
\lim_{b\rightarrow \infty} f_{Z_i}(b)b^2\log(b)=\infty
\eeqna
for  every  $i\in\{1,2\}$.
\end{theorem}
\IEEEproof see Section \ref{proof_theorem_exstans}.

Theorem \ref{theorem 3} asserts that if the ISR's PDF is tail heavy (as given  exactly in (\ref{25})), non pure-FS strategies are  possible and beneficial to both users. This condition is satisfied in important channel models including Rayleigh, Rician and Nakagami fading (as demonstrated in Section \ref{examples}). 

%


%

\section{The BIG in Common Channel Models}
\label{examples}
In this section we study the BIG in practical channel models such as Rayleigh, Nakagami and Rician. We will study the effect of different fading intensities   on the players' preferences,  the existence and uniqueness properties of  NE  points and  the performance gain.
\begin{figure}%
\centering
\subfigure[symmetric scenarios]{\label{Figure3a}\epsfig{figure=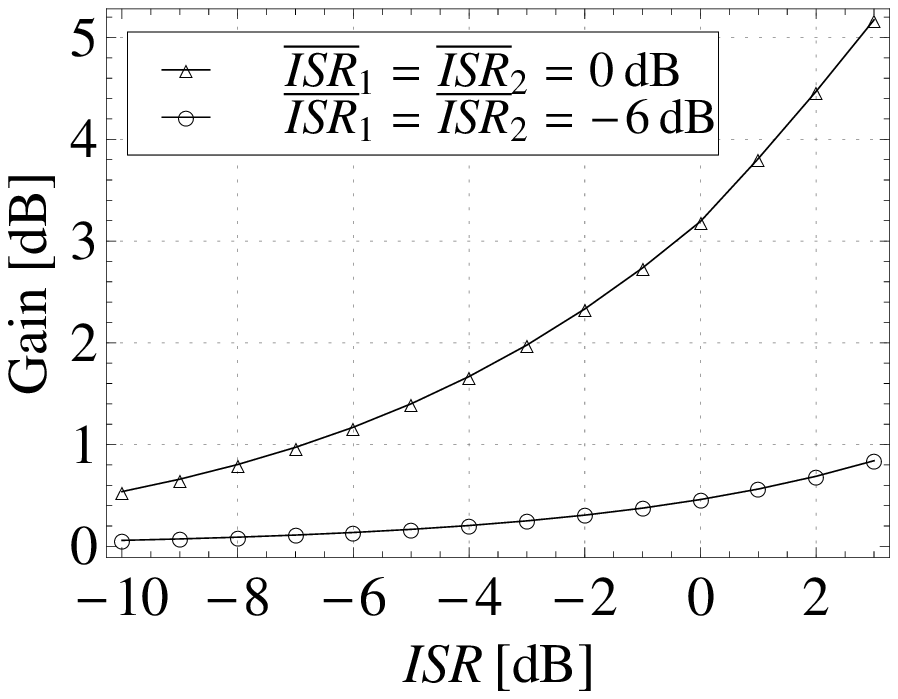,width=6.5cm}}
\qquad
\subfigure[weak strong scenario]{\label{Figure3b}\epsfig{figure=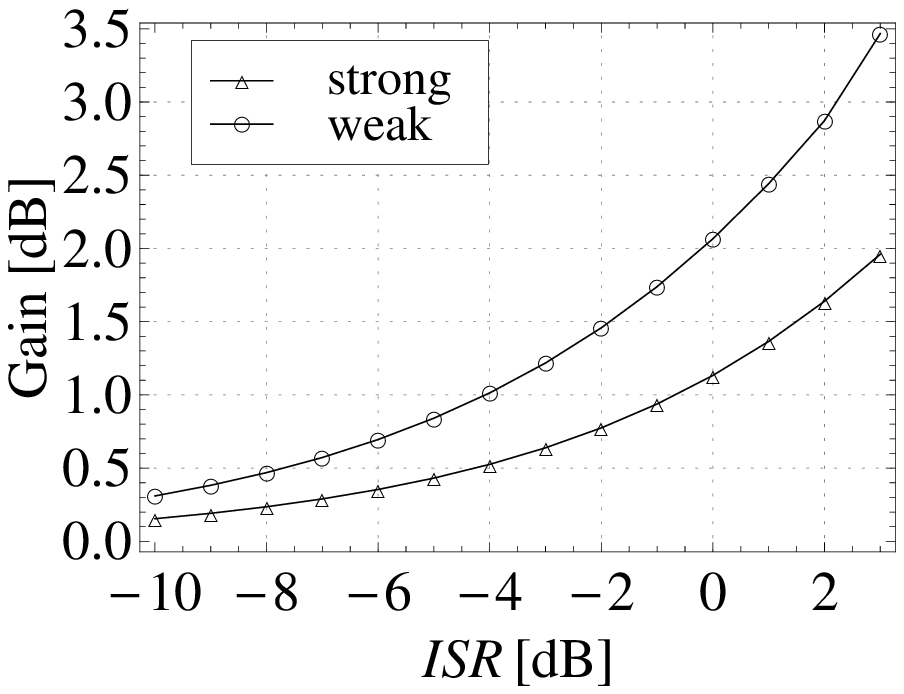,width=6.5cm}}
\caption{ The  difference (dB) between the conditional expected payoffs of non pure-FS  and  pure-FS  NE points as a function of $ISR$. The channel distributions are Rayleigh. (a) Two symmetric game scenarios: weak-weak (0 dB) and strong-strong (-6 dB). Each curve represents the gain in the corresponding scenario. (b) Weak-strong scenario, the weak $\aisr=-6$ dB whereas the strong  $\aisr=0$ dB.}
\label{Figure3}
\end{figure}

\subsection{Nakagami channel}

The Nakagami distribution \citep[see e.g.][Sec. 3.2.2]{goldsmith2005wc} is parameterized by averaged received  magnitude  and fading parameter  $m$, i.e.  $X$'s PDF is given by
\beq
f_X \left(x\right)=\left( \frac{m}{\overline{SNR}} \right)^m \frac{x^{m-1}}{\Gamma(m)}\exp\left( \frac{-mx}{\overline{SNR}}\right)
\eeq
where $\overline{SNR}$ is the averaged level of the SNR.

We now study the  existence of non pure-FS NE  points using Theorem \ref{theorem 3}. Denote the averages and the fading parameters of $X$ and $Y$ by $\overline{SNR},\;m_1$ and $\overline{INR},\;m_2$ respectively. Using the formula for transformation of random variables \citep[see e.g.][]{papoulis1965probability}, the PDF of $Z=X/Y$ is given by
\beqna
f_Z(z)= \frac{\aisr^{m_1}m_1^{m_1}m_2^{m_2}\Gamma(m_1+m_2)}{\Gamma(m_1)\Gamma(m_2)}\frac{ z^{m_2-1}}{\left(  m_2 z + \aisr m_1 \right )^{m_1+m_2} }
\eeqna
where $\aisr=\ainr/\asnr$.
Thus, by applying  Theorem \ref{theorem 3}, a sufficient condition for the existence of a non pure-FS NE point is that the fading coefficient of  the direct channel of both users must  satisfy
\beqna
\label{condition on m}
m_1\leq 1
\eeqna
In particular, this condition is satisfied in Rayleigh fading channels.

 Figure \ref{Figure3} shows the benefit of non pure-FS over pure-FS NE points for different values of $\aisr$ in Rayleigh fading channel (i.e. $m=1$ for  all paths).  Figure \ref{Figure3a} depicts a symmetric weak-weak scenario and a symmetric strong-strong scenario. In both  cases the conditional expected payoff is higher for  both players and  increases with the $ISR$. However, in the weak-weak scenario, the  gain is significant. Figure \ref{Figure3b} depicts a weak-strong scenario ($\aisr=-7$ corresponds to the strong).  In this case,  it is clear that the weak player gains more than the strong one, but non pure-FS is better for both of the players.

\begin{figure}
\centering
\epsfig{file=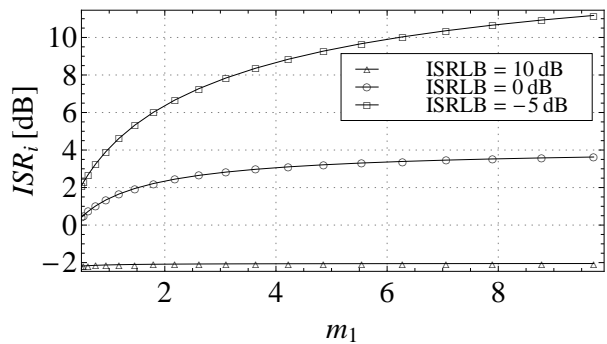,
width=0.5\textwidth} \caption{ Numerical evaluation of the  threshold ISR of the  approximate best response functions for Nakagami fading as a function of \(m_{1}\) - the  fading coefficient  in the direct channel. $\aisr=\ainr/\asnr$ stands for the ratio between the  averaged received INR and SNR. The value of $m_1$ is fixed and equal to 1 and the opponent's probability of  choosing FDM is  $a_j=0.2$. }
\label{Figurenakagami}
\end{figure}

In order to obtain insights into the BIG in Nakigimi channels, we address to numerical evaluation of the  approximate best response function (\ref{approximat best_response equation}) for different values of distribution parameters. To study the effect of $m_1$, the fading parameter in the direct channel,   Figure \ref{Figurenakagami} depicts  the threshold $ISR$ of the  approximate best response of player $i$ as a function of \(m_{1}\). This is evaluated for different values of $\overline{ISR}$. It is shown that the threshold $ISR$ is a decreasing function of $m_1$. This is violated only if interference  is very strong ($\overline{ISR}_i=10\;{\rm dB}$) whereas the threshold $ISR$ is hardly affected by the values of $m_1$. From this we deduce that a low fading effect (smaller probabilities of deep fade) in the direct channel (i.e. high values of $m_1$) encourages the use of FS (since the threshold $ISR$ increases).

\begin{figure}%
\centering
\qquad
\subfigure[]{\label{3figs_b}\epsfig{figure=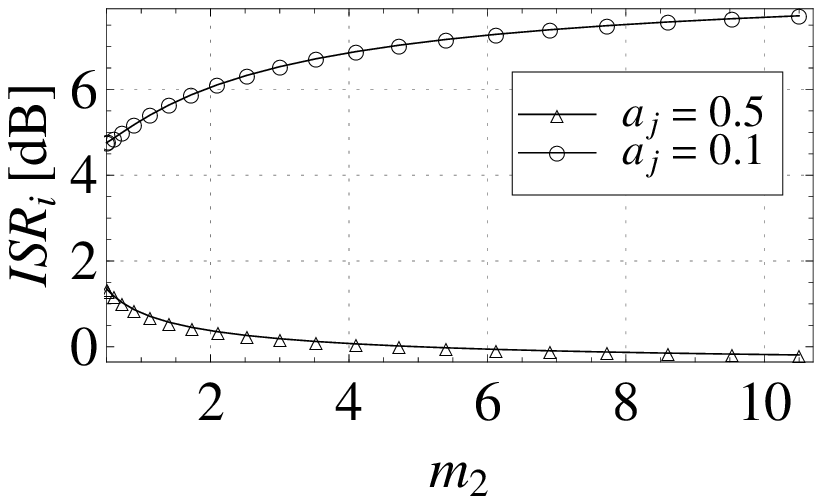,width=9cm}}
\\
\qquad
\subfigure[]{\label{3figs_c}\epsfig{figure=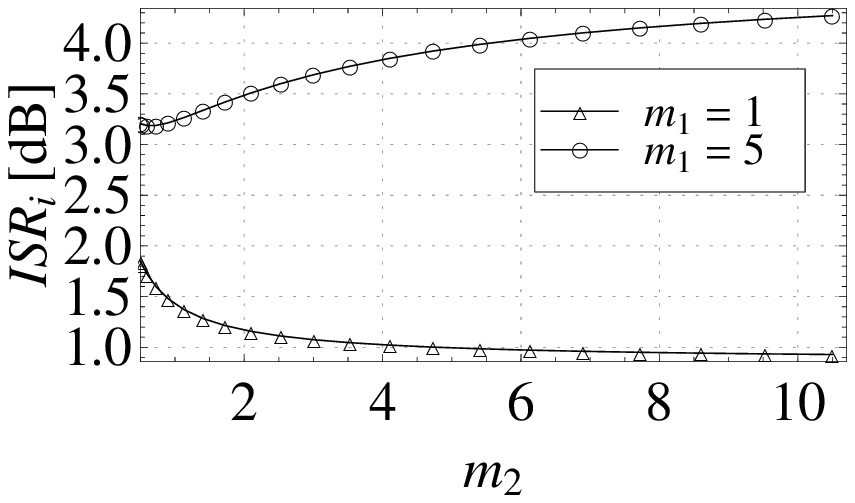,width=8.7cm}}
\caption{The threshold ISR  (above which player $i$ chooses FDM) as determined by the  approximate best response in a  Nakagami fading  channel. The horizontal  axes represent    the fading coefficient of the interference channel $m_2$. Figure  \ref{3figs_b} depicts the threshold ISR for both low and high level of $a_j$ (the opponent's probability of choosing FDM) with fixed  value of the fading parameter in the direct channel  $(m_1=1)$. Figure  \ref{3figs_c} depicts the threshold ISR for two levels of $m_{1}$  with   $a_j=0.1$.}
\label{6figs}
\end{figure}

In Figure \ref{6figs}, we study the effect of $m_2$, the fading parameter of the interfering channel. In this case we see that  the effect of \(m_{2}\) on the threshold ISR of the approximate best response depends on other factors such a \(m_{1}\) and \(a_{j}\). For low levels of \(a_{j}\), it can be seen in Figure \ref{3figs_b} that the threshold ISR is a increasing function of $m_2$ while it is a decreasing function for higher values on \(a_{j}\). In other words, if your (assuming that you are player $i$) opponent favours (does not favour) FDM, you should  consider FDM (FS) more strongly as the  interference  to your receiver becomes more dominant by  the line of sight path than by the reflected paths.
Figure \ref{3figs_c} shows  the same for the parameter  \(m_{1}\); i.e.  if  a player  experiences high probability of fading in the direct channel, he should  consider FDM (FS) more strongly if the  interference  to his receiver becomes more dominant by the line of sight than by the multipath. 

\begin{figure}%
\centering
\subfigure[Rayleigh fading]{\label{Nash_points_n1}\epsfig{figure=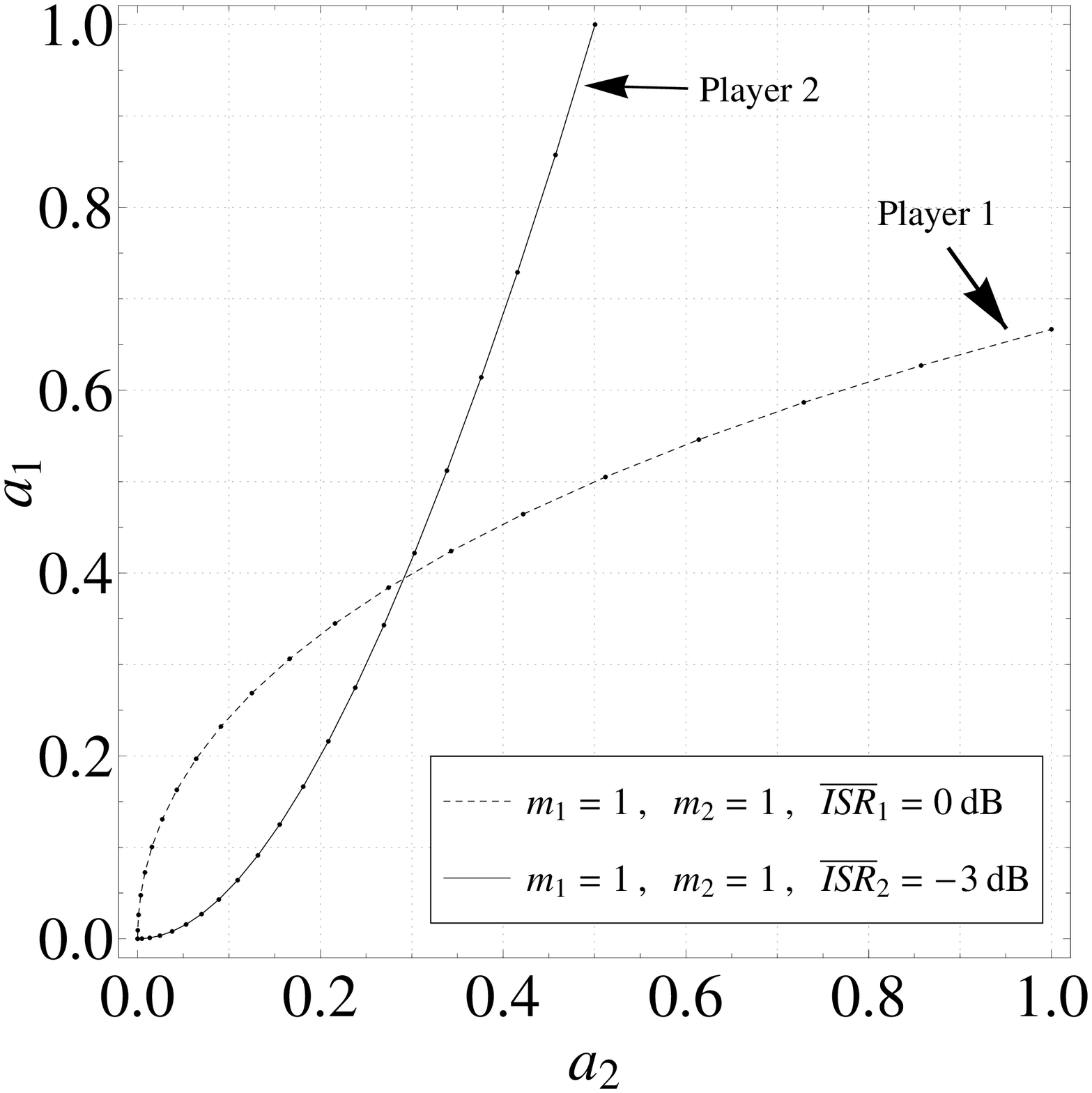,width=6.5cm}}
\qquad
\subfigure[Multiple equilibrium points]{\label{Nash_points_n2}\epsfig{figure=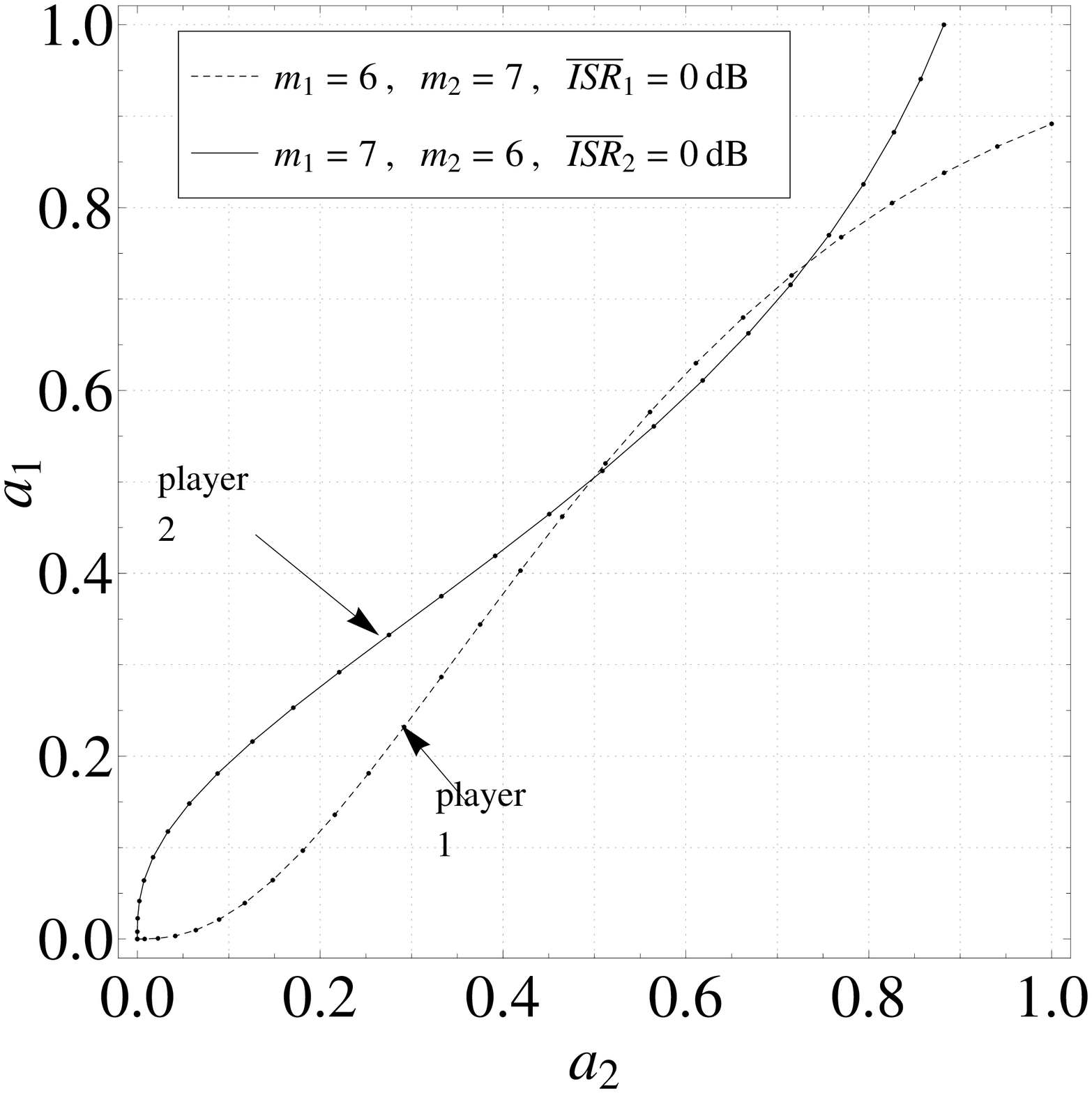,width=6.5cm}}
\caption{Numerical evaluation of the  $\epsilon$-near NE points  for Nakagami fading in different scenarios. The dashed (solid) lines represent player 1's (2's) best response for  given values of $m_1$ and $m_2$. Each intersection between dashed and solid lines is a $\epsilon$-near NE point. A user is considered "strong" ("weak") if its $\aisr$ his 10 dB (0 dB).}
\label{Nash_points_n11}
\end{figure}

\begin{figure}%
 \centering
{\label{Nash_points_n4}\epsfig{figure=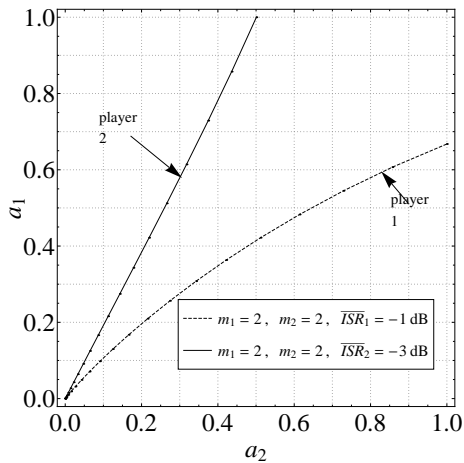,width=6.3cm}}
\caption{A scenario where the conditions of Theorem \ref{theorem 3} are not satisfied}
\label{Nash_points_n22}
\end{figure}

In Figures \ref{Nash_points_n11},\ref{Nash_points_n22} we study the existence properties of $\epsilon$-near NE points in different channel configurations. Figure \ref{Nash_points_n1}  shows a Rayleigh fading channel with two users and illustrates the $\epsilon$-near NE point. Figure \ref{Nash_points_n2} shows that  $\epsilon$-near NE
points are not necessarily unique. In Figure \ref{Nash_points_n22} we  show a scenario the conditions of  Theorem \ref{theorem 3} are not satisfied.


\section{Conclusions}

In this paper we  applied Bayesian games to analyze a two user wireless  interference channel  with incomplete information. Each player knows its  own direct and interfering channel magnitudes but only knows the statistics of its opponent's channel.

The main result of this paper is  the derivation of a non pure-FS $\epsilon$-NE point in the BIG with minimal coordination between users. This is a much better alternative than the pure-FS NE point which may be  very  inefficient. The non pure-FS point offers  better spectrum utilization efficiency  than the pure-FS Nash equilibrium. This is true for each user individually and in terms of  a global network. Through numerical examples, we demonstrated that this performance gain can be substantial. We further provided a sufficient condition for the existence of  non pure-FS $\epsilon$-NE and which is satisfied in particular in a Rayleigh fading channel. We also  demonstrated numerically  that  such points exist in many other scenarios.

In addition to the derivation of the non pure-FS NE points, in Section \ref{best_response_section} we presented the best response and the approximate best response function that converges in probability to the best response as the transmitted-power to noise ratio increases. The approximated best response  funcntion simply compares the measured interference-to-noise ratio to a threshold that depends on the opponents's probability of choosing FDM and on channel distribution. These results were later used in Section \ref{examples} to analyse selfish  and rational behaviour of wireless users as a function of the channel parameters. It was shown that:

\begin{itemize}
\item
 Strong fading (high probabilities for deep fade) in the direct channel encourages wireless selfish users to use FDM.
\item
 Strong fading in the interfering channel encourages selfish wireless users with strong fading in the direct channel to use FDM, while it has the opposite effect on users with weak fading in the direct channel.
\item
Strong fading in the interfering channel encourages selfish wireless users  to use FDM if the opponent chooses FDM with high probabilities, while it has the opposite effect if the opponent chooses FDM frequently.
\end{itemize}

\appendix
\subsection{Proof of Proposition  \ref{proposition for q}}
\label{appendix_a}
 Observe that   $r(a,b)$ is a continuous, differentiable and strictly increasing function of $b$ for every $a$. It can be shown that   $r(a,1/2)<0$ and that $\lim_{b\longrightarrow \infty}r(a,b)>0$ for all $a>0$. Thus, $r(a,q(a))=0$ defines an implicit differentiable function $q(a)$ that satisfies  $q(a)>1/2$ for every $0< a\leq1$.

We show that  $q(a)$ is a strictly monotonic decreasing function of $a$.  This can be established by observing the derivative of  $q(a)$
\beqna
\barrayl
\label{19}
q'(a)=& \left(-\frac{q(a)\left( 1 + q(a) \right)
      \left( 2 + q(a) \right)
      \left( 1 + 2\,q(a) \right)
      }{2 + 4q(a) -
      a\,q(a) + a{q^2(a)}} \right) \left(2+2\log (1 +  q(a))- \log (2 + q(a))  -
        \log (1 + 2\,q(a)) \right)
        \earray
\eeqna
Since $q(a)>1/2$ the derivative is negative.
\hfill $\Box$

\subsection{Proof of Lemma \ref{equivalence_condition}}
\label{equivalence_condition proof}

Since $\check S_i$ and $\tilde S_i$ are binaries in their range, it is sufficient to show that
\beqna
\vert P((X_i,Y_i)\in \check{D}^{a_j}_i)-P((X_i,Y_i)\in\tilde D_i^{a_j})\vert \leq \epsilon,\; \;\forall\; \bar p>\bar p_0
\eeqna
Henceforth, the indices $i,j$ are omitted, $a$ will denote $a_j$ and $\check{D}^{a},\; \tilde D^{a}$ will denote $\check{D}^{a_j}_i,\; \tilde D_i^{a_j}$.

Let $\bar p_n,\; \alpha_n$ be  sequences satisfying $\lim_{n\rightarrow \infty} \bar p_n,\alpha_n=\infty$  such that \(\alpha_n=o\left( \bar p_n \right)\),\footnote{For  deterministic sequences \(\alpha_n,\beta_n \) with $\lim_{n\rightarrow \infty }\alpha_n/\beta_n=M$ we say that $\alpha_n=O(\beta_n) $ if \(M\) is finite and non zero and $\alpha_n=o(\beta_n)$ if \(M=0\).}  denote $X^n=\bar p_n\vert H_{ii}\vert^2/\sigma_N^2$, $
Y^n=\bar p_n\vert H_{ij}\vert^2/\sigma_N^2$, $P_n(A)=$$P\left( ( X^n,Y^{n})\in A\right)$. Further denote ${\cal A}_{n}=\{X^{n}>\alpha_n\}$ and ${\cal B}_{n}=\{Y^{n}>\alpha_n/2\}$.

\begin{figure}
\centering
\epsfig{figure=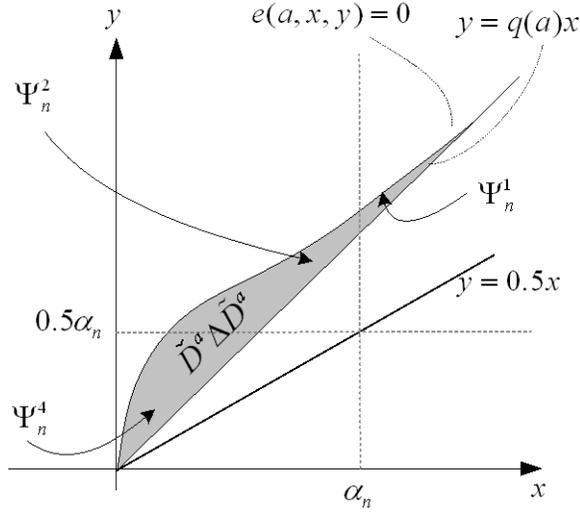}
\caption{Graphic illustration of  the partition in the proof of Lemma \ref{equivalence_condition}.}
\label{proof_intuition}
\end{figure}

 Define \beqna\begin{array}{l}
\Psi _{n}^i=\left(\check{D}^{a}\Delta {{{\tilde{D}}}^{a}} \right)\bigcap{G_i}
\end{array}
 \eeqna
 (see Figure \ref{proof_intuition} for illustration) where $G_1={\cal A}_{n}\bigcap {\cal B}_{n}$, $G_2={\cal A}_{n}^c\bigcap {\cal B}_{n}$, $G_3={\cal A}_{n}\bigcap {\cal B}_{n}^c$ and $G_4={\cal A}_{n}^c\bigcap {\cal B}_{n}^c$. This partition satisfies
\beq\label{20_85}
   {{P}_{n}}(\check{D}^{a}\Delta \tilde D^a) =\sum_{q=1}^4P_{n}\left(\Psi^q_n\right)
\eeq
and \beqna\label{20_86}
 P_n\left(\Psi^3_n\right)&{}={}&0
 \\P_n\left(\Psi^2_n\right)&{}\leq{}&F_{\vert H_{ii}\vert ^2}\left(\sigma_N^2 \frac {\alpha_n} {\bar p_n}\right)
 \\\label{20_87}
P_n\left(\Psi^4_n\right)&{}\leq{}& F_{\vert H_{ii}\vert ^2}\left(\sigma_N^2 \frac {\alpha_n} {\bar p_n}\right)F_{\vert H_{ij}\vert ^2}\left(\sigma_N^2 \frac {\alpha_n} {2\bar p_n}\right)
\eeqna
where (\ref{20_86}) is true because both strategies are identical if $y\leq 0.5 x$.
Therefore, to show the first part of the Lemma (Equation \eqref{decreases_to_zero})   , it is sufficient to show that $P_n\left(\Psi^1_n\right)=o(1)$. This follows from the fact that for every $a>0$,
\beqna
\lim_{n\rightarrow \infty} e(\alpha_n,q(a)\alpha_n,a)-\hat e(\alpha_n,q\alpha_n,a)=0
\eeqna
Thus
\beqna
\lim\sup_n \Psi^1_n = \phi
\eeqna
and from the continuity from above of measures \citep[see e.g.][Theorem 1.8]{Folland} it follows that
\beq
\lim_{n\rightarrow \infty} P_n\left(\Psi_n\right)=0
\eeq
which establishes the first part of the Lemma.

For the second part of the Lemma, we will show that
\begin{eqnarray}
P_n(\check{D}^{a} \Delta \tilde D^{a})\leq O\left(\ \frac{\sigma_N^2}{\bar p_n^{}}+\left(F_{\vert H_{ii}\vert ^2}\left(\sigma_N^2 \frac {\alpha^2_n} {\bar p_n}\right)\right)^{2}+\prod_{q=1}^2\left(F_{\vert H_{iq}\vert ^2}\left( \frac {\sigma^2_N}{\bar p_{n}^{1-\nu} }\right)\right)\right)
\label{p20_77}\end{eqnarray} This requires an additional analysis of  $P_n(\Psi_n^1)$ and $P_n(\Psi_n^2)$. For the term $P_n\left(\Psi^1_n\right)$, we first assume that that $Y^{n}>q(a)X^{n}$. In this case
\beqna
P_n\left(\left.\Psi _{n}^1\right \vert Z>q(a),{\cal A}_{n}\right)=P\left(\right.e(X^n,Y^{n},a)<0 \left\vert {\cal A}_{n},Y^{n}>q(a)X^n  \right)=P^1_n+P^2_n
\eeqna
where
\beqna
\begin{array}{ll}
P^1_n=P\left(\left. e(X^n,Y^{n},a)<0, q(a) < Z < q(a) + \frac{1}{\gamma _n}\right\vert {\cal A}_{n},Y^{n}>q(a)X^n  \right)
\end{array}
\eeqna
\beqna
\begin{array}{ll}
P^2_{n}=P\left( e(X^n,Y^{n},a)<0,q(a)+ \frac{1}{\gamma _n} \leq  Z \left\vert {\cal A}_{n},Y^{n}>q(a)X^n   \right. \right)
\end{array}
\eeqna
and  $\gamma_n=o(\alpha_n)$.  Before evaluating $P^1_n$ and $P^2_{n}$ the function   $e(x,y,a)$ will be simplified by   substituting  $y=zx $ (which is possible because $x,y>0$) \beqna\begin{array}{ll}
   e(x,z\,x,a)= & \frac{1}{2}a\log (x+1)-a\left( \frac{1}{2}\log \left( \frac{x}{2(xz+1)}+1 \right)+\frac{1}{2}\log \left( \frac{x}{2}+1 \right) \right)   \\
   {} & -(1-a)\log \left( \frac{x}{xz+2}+1 \right)+\frac{1}{2}(1-a)\log \left( \frac{2x}{xz+2}+1 \right)\end{array}
\eeqna
which is a bounded function of $z$ for every $x$. Furthermore,  it is easy to verify that the function
\(e(\frac{1}{w},\frac{z}{w},a)\) is infinitely differentiable with bounded derivatives at $w=0$. Thus, since   \beqna
\begin{array}{ll}
t(z,a)\triangleq \lim_{w\rightarrow 0}e(\frac{1}{w},\frac{z}{w},a)=\frac{1}{2} \left(a \left(1-\log \left(\frac{1}{2 z}+1\right)\right)\right.\\ \left.+(1-a) \log
   \left(1+\frac{2}{z}\right)+2 (a-1) \log \left(\frac{1}{z}+1\right)\right)
   \end{array}
\eeqna
is bounded, continuous and differentiable on $z>0.5$ for every \(a\) it is  possible to expand $e(x,z\,x,a)$ with respect to $1/x$  and obtain \beqna\label{taylor expansion of e}
e(x,z\,x,a)=r(a,z)+Q(a,z)\frac{1}{x}+O {{\left( \frac{1}{x^2} \right)}}
\eeqna
where
$r(a,z)$ is defined in \eqref{17cc}, the residual absolute value can be bounded by $M/x^2$ where $M$ is finite and \beqna
Q(a,z)=\frac{{ - 2a{z^4} - 7a{z^3} - 6a{z^2} - 7az - 2a + 8z + 4}}{{2z(z + 1)(z + 2)(2z + 1)}}
\eeqna
is  bounded  for every $z,a$;
 furthermore, since $r(a,z)$ is a continuous and increasing function of $z$ for every $a>0$ (as shown in Proposition \ref{proposition for q}) that satisfies $r(a,q(a))=0$, it follows that \beqna
r(a,z) =R(a)(z - q(a)) + O\left( {{{(z - q(a))}^2}} \right)
\eeqna
where the
 residual absolute value can be bounded by $M/(z - q(a))^2$ where $M$ is finite for every $a>0$ and \beqna
R(a) = \frac{{\left( {a{q(a)^2} - aq(a) + 4q(a) + 2} \right)}}{{2q(a)(q(a) + 1)(q(a) + 2)(2q(a) + 1)}}
\eeqna
 is bounded and positive for every $0\leq a\leq 1$  because $q(a)>1/2$.

In what follows it is shown that for sufficiently large $n$, $P^2_{n}=0$. To see this, observe that for every $z\geq (a) + \frac{1}{\gamma _n}$ \beqna
\begin{array}{lll}
 e(x,z\,x,a)&\geq r(a,z)-\frac{M_1}{x}- \frac{M_2}{x^2} \geq r(a,z)-\frac{M_1}{\alpha_n}- \frac{M_2}{\alpha_n^2}\\ &\geq\frac{R(a)}{\gamma_n} + O\left( \frac{1}{\gamma_n^2} \right)-\frac{M_1}{\alpha_n}- \frac{M_2}{\alpha_n^2}
 \end{array}
 \eeqna
 where \(M_{1}\), \(M_{2}\)  are positive and finite for every $z$ and $a$. Therefore,
 \beqna
 \gamma_n e(x,z x,a)\geq{R(a)} + O\left( \frac{1}{\gamma_n}
  \right)-\frac{M_1\gamma_n}{\alpha_n}- \frac{M_3\gamma_n }{\alpha_n^2}
 \eeqna
 and becomes  the \(R(a)>0,\forall a>0\) and because $M_1$ and $M_2$ are bounded, it follows that \(P^{2}_n=0\) for sufficiently large \(n\).

  It remains to show that $P^1_{n}$ decreases like $1/\bar p$.
By substituting  the series expansions of $r(a,z)$ into (\ref{taylor expansion of e})
it follows that for every $z\in( q(a) , q(a) + 1/{\gamma _n}),\; x>\alpha_n$
\beqna\barrayl e(x,y,a)=
R(a)(z - q) + \frac{{Q(a,z)}}
{x} + O\left( {\frac{1}
{{{x^2}}}} \right) + O{(z - q)^2}\\
 ~~\leq R(a)\left( {z - q(a)} \right) - \frac{{{M_1}}}
{x} - \frac{{{M_2}}}
{{{x^2}}} - {M_3}{(z - q)^2} \leq {\eta _n}\left( {z - q(a)} \right) - \frac{{{\xi _n}}}
{x}
\earray\eeqna
where $\eta_n=R(a)-{{M_3}}/{\gamma_n}$ and $\xi_n= M_{1} + {{{M_2}}}/
{{{\alpha_n}}}$.

Thus,
 \begin{eqnarray}
 \begin{array}{ll}
   P_{n}^{1}&\le \frac{P\left( 0<{{Y}^{n}}-q(a){{X}^{n}}<\min ({{\xi }_{n}}/{{\eta }_{n}},{{X}^{n}}/{{\gamma }_{n}}),{\cal A}_{n} \right)}{P\left( {{Y}^{n}}>{{X}^{n}}q(a),{\cal A}_{n} \right)} =\frac{\int\limits_{{{\alpha }_{n}}}^{\infty }{\int\limits_{xq(a)}^{q(a)x+{{\xi }_{n}}/{{\eta }_{n}}}{{{f}_{{{Y}^{n}}}}(y){{f}_{{{X}^{n}}}}(x)dydx}}}{\int\limits_{{{\alpha }_{n}}}^{\infty }{\int\limits_{xq(a)}^{\infty }{{{f}_{{{Y}^{n}}}}(y){{f}_{{{X}^{n}}}}(x)dydx}}}\\&=\frac{\int\limits_{{{\alpha }_{n}}}^{\infty }{\left( {{F}_{{{Y}^{n}}}}\left( q(a)x+{{\xi }_{n}}/{{\eta }_{n}} \right)-{{F}_{{{Y}^{n}}}}\left( q(a)x \right) \right){{f}_{{{X}^{n}}}}(x)dx}}{\int\limits_{{{\alpha }_{n}}}^{\infty }{\left( 1-{{F}_{{{Y}^{n}}}}\left( q(a)x \right) \right){{f}_{X}}(x)dx}}=\frac{{{\mu }_{n}}}{{{\lambda }_{n}}}
\end{array}
 \end{eqnarray}

Note that
 $\lim \inf_n\lambda_n$ \textgreater 0, because

\beqna\begin{array}{lll} \lambda_{n}&{}={}&
\int\limits_{\alpha _n}^\infty  \left( 1 - F_{\left| {H_{ij}} \right|}^2\left( \frac{\sigma _v^2q(a)x}{{\bar p}_n} \right) \right){f_{X^n}}(x)dx
\ge \int\limits_{\alpha _n}^{{{\bar p}_n}} \left( {1 - F_{{{\left| {{H_{ij}}} \right|}^2}}\left( {\frac{{\sigma _v^2q(a)x}}{{{{\bar p}_n}}}} \right)} \right){f_{X^n}}(x)dx
 \\ &{}\geq{}&
 \int\limits_{{\alpha _n}}^{{{\bar p}_n}} {\left( {1 - {F_{{{\left| {{H_{ij}}} \right|}^2}}}\left( {\sigma _v^2q(a)} \right)} \right){f_{X^n}}(x)dx}
= \left( {1 - {F_{{{\left| {{H_{ij}}} \right|}^2}}}\left( {\sigma _v^2q(a)} \right)} \right)\times\left( {F_{{{\left| {{H_{ij}}} \right|}^2}}}\left( {\sigma _v^2q(a)} \right)\right.\\&{}{}&\left. - {F_{{{\left| {{H_{ij}}} \right|}^2}}}\left( {\frac{{\sigma _v^2q(a){\alpha _n}}}{{{{\bar p}_n}}}} \right) \right)\xrightarrow[n\rightarrow\infty]{} \left( {1 - {F_{{{\left| {{H_{ij}}} \right|}^2}}}\left( {\sigma _v^2q(a)} \right)} \right){F_{{{\left| {{H_{ij}}} \right|}^2}}}\left( {\sigma _v^2q(a)} \right) > 0,
\end{array}\eeqna
therefore, the first term of $P^1_{n}$ decreases like $\mu_n$
\beqna\begin{array}{ll}
\mu_n&\leq\int\limits_{{0}}^\infty  {\left( {{F_{{{\left| {{H_{ij}}} \right|}^2}}}\left( {\frac{{\sigma _v^2(q(a)x + \xi_n/{\eta _n})}}
{{{{\bar p}_n}}}} \right) - {F_{{{\left| {{H_{ij}}} \right|}^2}}}\left( {\frac{{\sigma _v^2q(a)x}}
{{{{\bar p}_n}}}} \right)} \right){\frac{{\sigma _v^2}}
{{{{\bar p}_n}}}{f_{{{\left| {{H_{ii}}} \right|}^2}}}\left( {\frac{{\sigma _v^2x}}
{{{{\bar p}_n}}}} \right)}
}dx\\ \label{prodoct to use for the dominant convergence theorem}&=\frac{\sigma_v^2\xi_n}
{\eta _n\bar p_n}\int\limits_0^\infty  {\frac{{{F_{{{| {{H_{ij}}} |}^2}}}\left( {q(a)v +{\sigma_v^2}\xi_n/(
{\eta _n}{\bar p}_n}) \right) - {F_{{{| {{H_{ij}}} |}^2}}}\left( {q(a)v} \right)} }{ {\sigma_v^2}\xi_n/(
{\eta _n}{\bar p}_n)  }{f_{{{\left| {{H_{ii}}} \right|}^2}}}\left( v \right)dv}
\end{array}
 \eeqna

Recall that by hypothesis $f_{\left| H_{iq} \right|^2}\left( v \right)$ is bounded for every \(v>0\). Thus  by the LaGrange  mean value theorem, for every \(v\geq\delta\)
\beqna\barrayl
\frac{{{F_{{{| {{H_{ij}}} |}^2}}}\left( {q(a)v +{\sigma_v^2}\xi_n/
{\eta _n}\bar p_n)} \right) - {F_{{{| {{H_{ij}}} |}^2}}}\left( {q(a)v} \right)} }{{\sigma_v^2}\xi_n/(
{\eta _n}{\bar p}_n)}\leq\sup_{\theta\in [0,1]}\left({f_{{{\left| {{H_{ij}}} \right|}^2}}}\left( {q(a)v}+ \frac{\theta\xi_n\sigma_v^2}
{{{\eta _n}{{\bar p}_n}}} \right)\right)\leq M
\earray\eeqna by invoking the dominant convergence theorem \citep[see e.g.][Theorem 2.24]{Folland} on the integral in (\ref{prodoct to use for the dominant convergence theorem})
\beqna\begin{array}{lll}\lim_{\delta\rightarrow 0}&
\lim_{n\rightarrow \infty}\int\limits_\delta^\infty  {\frac{{{F_{{{| {{H_{ij}}}|}^2}}}\left( {q(a)v +  {\sigma_v^2}\xi_n/(
{\eta _n}\bar p_n)   } \right) - {F_{{{| {{H_{ij}}} |}^2}}}\left( {q(a)v} \right)} }{{\sigma_v^2}\xi_n/(
{\eta _n}\bar p_n)}{f_{{{\left| {{H_{ii}}} \right|}^2}}}\left( v \right)dv}\\&=\lim_{\delta\rightarrow 0}
\int\limits_\delta^\infty  {{f_{{{\left| {{H_{ij}}} \right|}^2}}}\left( {q(a)v} \right)}{f_{{{\left| {{H_{ii}}} \right|}^2}}}\left( v \right)dv=
\int\limits_0^\infty  {{f_{{{\left| {{H_{ij}}} \right|}^2}}}\left( {q(a)v} \right)}{f_{{{\left| {{H_{ii}}} \right|}^2}}}\left( v \right)dv\end{array}
\label{22_106_}\eeqna
where \eqref{22_106_} is true  because ${f_{{{\left| {{H_{iq}}} \right|}^2}}}\left(v \right),~i,q\in\{1,2\}$ are probability densities. Furthermore, it is positive and finite for every $a$. From this it follows that
\beqna\label{limis} P_n^1\leq  O\left( \frac{\sigma_v^2}
{{{}{{\bar p}_n}}}\right)
\eeqna and therefore \beqna
\label{p given Z>q}
P_n\left(\left.\Psi _{n}^1\right/Z>q(a),{\cal A}_{n}\right)\leq  O\left(\frac{1}{\bar p_n}\right)
\eeqna

We now assume that that $Y^{n}\leq q(a)X^{n}$. In this case
\beqna
P_n\left(\left.\Psi _{n}^1\right \vert Z\leq q(a),{\cal A}_{n}\right)=P\left(\right.e(X^n,Y^{n},a)>0 \left\vert {\cal A}_{n},Y^{n}\leq q(a)X^n  \right)=P^1_n+P^2_n
\eeqna
where
\beqna
\begin{array}{ll}
P^1_n=P\left(\left. e(X^n,Y^{n},a)\geq 0, q(a)- \frac{1}{\gamma _n} < Z < q(a) \right\vert {\cal A}_{n},Y^{n}\leq q(a)X^n  \right)
\end{array}
\eeqna
\beqna
\begin{array}{ll}
P^2_{n}=P\left(\left. e(X^n,Y^{n},a)\geq 0, Z \leq q(a)-\frac{1}{\gamma _n}  \right \vert {\cal A}_{n},Y^{n}\leq q(a)X^n   \right)
\end{array}
\eeqna

In what follows it is shown that for sufficiently large $n$, $P^2_{n}=0$. To see this, observe that for every $z\leq q (a) -\frac{1}{\gamma _n}$ \beqna
 e(x,z\,x,a)\leq r(a,z)+\frac{M_1}{x}+ \frac{M_2}{x^2} \leq r(a,z)+\frac{M_1}{\alpha_n}+ \frac{M_2}{\alpha_n^2}\\ \leq-\frac{R(a)}{\gamma_n} + O\left( \frac{1}{\gamma_n^2} \right)+\frac{M_1}{\alpha_n}+ \frac{M_2}{\alpha_n^2}
 \eeqna
 where \(M_{1}\), \(M_{2}\)  are positive and finite for every $z$ and $a$. Therefore,
 \beqna
 \gamma_n e(x,z x,a)\leq{-R(a)} + O\left( \frac{1}{\gamma_n}
  \right)+\frac{M_1\gamma_n}{\alpha_n}+ \frac{M_3\gamma_n }{\alpha_n^2}
 \eeqna
 and become  the \(R(a)>0,\forall a>0\) and because $M_1$ and $M_2$ are bounded, it follows that \(P^{2}_n=0\) for sufficiently large \(n\).

  It remains to show that $P^1_{n}$ decreases like $1/\bar p$.
By substituting  the series expansions of $r(a,z)$ into (\ref{taylor expansion of e})
it follows that for every $z\in( q(a)-1/{\gamma _n} , q(a)),\; x>\alpha_n$
\beqna\barrayl e(x,y,a)\leq R(a)\left( {z - q(a)} \right) + \frac{{{M_1}}}
{x} + \frac{{{M_2}}}
{{{x^2}}} + {M_3}{(z - q)^2} \leq {\eta _n}\left( {z - q(a)} \right) + \frac{{{\xi _n}}}
{x}
\earray\eeqna
where $\eta_n=R(a)+{{M_3}}/{\gamma_n}$ and $\xi_n= M_{1} + {{{M_2}}}/
{{{\alpha_n}}}$.
Thus,
 \begin{eqnarray}
 \begin{array}{ll}
   P_{n}^{1}&\le \frac{P\left(-\min ({{\xi }_{n}}/{{\eta }_{n}},{{X}^{n}}/{{\gamma }_{n}})<{{Y}^{n}}-q(a){{X}^{n}}<0,{\cal A}_{n} \right)}{P\left( {{Y}^{n}}\leq{{X}^{n}}q(a),{\cal A}_{n} \right)} \\&=\frac{\int\limits_{{{\alpha }_{n}}}^{\infty }{\left( {{F}_{{{Y}^{n}}}}\left( q(a)x \right)-{{F}_{{{Y}^{n}}}}\left( q(a)x-{{\xi }_{n}}/{{\eta }_{n}} \right) \right){{f}_{{{X}^{n}}}}(x)dx}}{\int\limits_{{{\alpha }_{n}}}^{\infty }{ {{F}_{{{Y}^{n}}}}\left( q(a)x  \right){{f}_{X}}(x)dx}}=\frac{{{\mu }_{n}}}{{{\lambda }_{n}}}
\end{array}
 \end{eqnarray}
Note that
 $\lim \inf_n\lambda_n$ \textgreater 0, to see this
\beqna\begin{array}{lll} \lambda_{n}&{}={}&
\int\limits_{\alpha _n}^\infty  F_{\left| {H_{ij}} \right|}^2\left( \frac{\sigma _v^2q(a)x}{{\bar p}_n} \right) {f_{X^n}}(x)dx
\ge \int\limits_{\bar p _n}^{\infty}  {F_{{{\left| {{H_{ij}}} \right|}^2}}\left( {\frac{{\sigma _v^2q(a)x}}{{{{\bar p}_n}}}} \right)} {f_{X^n}}(x)dx
 \\ &{}\geq{}&
 \int\limits_{\bar p _n}^\infty { { {F_{{{\left| {{H_{ij}}} \right|}^2}}}\left( {\sigma _v^2q(a)} \right)} {f_{X^n}}(x)dx}
=  { {F_{{{\left| {{H_{ij}}} \right|}^2}}}\left( {\sigma _v^2q(a)} \right)} \times\left( {1-F_{{{\left| {{H_{ij}}} \right|}^2}}}\left( {\sigma _v^2q(a)} \right)\right)
\end{array}\eeqna
therefore, the first term of $P^1_{n}$ decreases like $\mu_n$
\beqna\begin{array}{ll}
\mu_n&\leq\int\limits_{{0}}^\infty  {\left( {{F_{{{\left| {{H_{ij}}} \right|}^2}}}\left( {\frac{{\sigma _v^2(q(a)x)}}
{{{{\bar p}_n}}}} \right) - {F_{{{\left| {{H_{ij}}} \right|}^2}}}\left( {\frac{{\sigma _v^2(q(a)x- \xi_n/{\eta _n}})}
{{{{\bar p}_n}}}} \right)} \right){\frac{{\sigma _v^2}}
{{{{\bar p}_n}}}{f_{{{\left| {{H_{ii}}} \right|}^2}}}\left( {\frac{{\sigma _v^2x}}
{{{{\bar p}_n}}}} \right)}
}dx\\ \label{prodoct to use for the dominant convergence theorem2}&=\frac{\sigma_v^2\xi_n}
{\eta _n\bar p_n}\int\limits_0^\infty  {\frac{{F_{{{| {{H_{ij}}} |}^2}}}\left( {q(a)v} \right)-{{F_{{{| {{H_{ij}}} |}^2}}}\left( {q(a)v -{\sigma_v^2}\xi_n/(
{\eta _n}{\bar p}_n}) \right)  } }{ {\sigma_v^2}\xi_n/(
{\eta _n}{\bar p}_n)  }{f_{{{| {{H_{ii}}} |}^2}}}\left( v \right)dv}\leq  O\left( \frac{\sigma_v^2}
{{{}{{\bar p}_n}}}\right)
\end{array}
 \eeqna which leads to
 \beqna
\label{p given Z>q}
P_n\left(\left.\Psi _{n}^1\right/Z\leq q(a),{\cal A}_{n}\right)\leq  O\left(\frac{1}{\bar p_n}\right)
\eeqna
and therefore \beqna\label{babababa}P_n(\Psi _{n}^1
)\leq O\left( 1/\bar p_n\right)\eeqna

It remains to evaluate $P_{n}\left(\Psi _{n}^2\right)$. Note that  \beqna
P_n\left(\Psi _{n}^2\right)=P\left( {\cal B}_{n},{\cal A}_{n}^c\right)P_{n}\left(\left.\Psi _{n}^2 \right\vert {\cal B}_{n},{\cal A}_{n}^c\right)
\eeqna
 and
\beqna
P\left( {\cal B}_{n},{\cal A}_{n}^c\right)\leq O\left( {F_{{{\left| {{H_{ii}}} \right|}^2}}}\left( {\frac{{\sigma _v^2{\alpha _n}}}
{{{{\bar p}_n}}}} \right)\right)
\eeqna
it follows that
\beqna
P_n\left(\Psi _{n}^2\right)\leq P_{n}\left(\left.\Psi _{n}^2 \right\vert  {\cal C}_n\right)O\left( {F_{{{\left| {{H_{ii}}} \right|}^2}}}\left( {\frac{{\sigma _v^2{\alpha _n}}}
{{{{\bar p}_n}}}} \right)\right)
\label{page23 eq 122}\eeqna
where \(\mathcal{C}_n\) = $\mathcal{A}_n^c\bigcap {\cal B}_n$. Furthermore
\beqna\barrayl
  P_{n}\left( \left. {\Psi _n^2} \right| \mathcal{C}_n\right)&=&
  P_{n}\left( {\left. {\Psi _n^2} \right|\mathcal{C}_n,Z > {\alpha _n}/2} \right) P\left( {\left. {Z > {\alpha _n}/2} \right|\mathcal{C}_n} \right) +P_{n}\left( {\left. {\Psi _n^2} \right|\mathcal{C}_n,Z \leq {\alpha _n}/2} \right) P\left( {\left. {Z \leq {\alpha _n}/2} \right|\mathcal{C}_n} \right)\\
&\leq& P_{n}\left( {\left. {\Psi _n^2} \right|\mathcal{C}_n,Z > {\alpha _n}/2} \right) P\left( {\left. {Z > {\alpha _n}/2} \right|\mathcal{C}_n} \right)
+ {F_{{{\left| {{H_{ij}}} \right|}^2}}}\left( {\frac{{\sigma _v^2\alpha _n^2}}
{{{{\bar p}_n}}}} \right)\earray\label{23_124}
\eeqna
where the last inequality is due to
\beqna
\begin{array}{lll}
\label{page23 eq126}  {
P\left( {\left. {Z \leq {\alpha _n}/2} \right|\mathcal{C}_n} \right) = \frac{{P\left( {{\alpha _n}/2 < Y^{n} < {X^n}{\alpha _n}/2,\,{X^n} \leq {\alpha _n}} \right)}}
{{P\left( {Y^{n} > {\alpha _n}/2,\,{X^n} \leq {\alpha _n}} \right)}}}\\ ~~~  \leq \frac{{P\left( {{\alpha _n}/2 < Y^{n} < {\alpha ^2}_n/2,\,{X^n} \leq {\alpha _n}} \right)}}
{{P\left( {Y^{n} > {\alpha _n}/2,\,{X^n} \leq {\alpha _n}} \right)}} = P\left( {{\alpha _n}/2 < Y^{n} < {\alpha ^2}_n/2} \right) \leq {F_{{{\left| {{H_{ij}}} \right|}^2}}}\left( {\frac{{\sigma _v^2\alpha _n^2}}
{{{{\bar p}_n}}}} \right)
\end{array}
\eeqna
It remains to calculate the term \beqna
P\left( {\left. {\Psi _n^2} \right\vert \mathcal{C}_n,Z > {\alpha _n}/2} \right)= P\left( {\left. {e\left( {{X^n},Y^{n},a} \right) < 0} \right|\mathcal{C}_{n},Y^{n}/{X^n} > {\alpha _n}/2} \right)
\label{probabilty of 126}\eeqna
To evaluate \eqref{probabilty of 126}, consider the function \(e\left( {x,y,a} \right)-T(x)\) where \(T(x)=\frac{a}{2} \log \left(1+{2x}/({x+2})\right) \). Similar to the derivation of (\ref{taylor expansion of e}) we obtain
\begin{eqnarray} e\left( {\frac{y}{z},y,a} \right) -T\left( \frac{y}{z} \right)=r\left(  a,z\right)+\frac{a z^2-5 a z-2 a+8
   z+4}{2(z+1) (z+2) (2 z+1)}\frac{1}{y}+O\left( \frac{1}{y^2} \right)\end{eqnarray}and because \(r\left (  a,z\right )\) is an increasing and positive  function of \(z\) for   \(z>q(a) \) and for every  \textit{a~}  and because \(T\left( {y}/{z} \right)\geq0\) for every $y,z\geq0 $,   the RHS of \eqref{probabilty of 126} is equal to zero for  sufficiently large $n$.
Thus, by combining \eqref{page23 eq126} and \eqref{page23 eq 122}, it follows that \begin{equation}\label{23_131}
 P\left(\left( X^n,Y^n \right)\in\Psi _{n}^2\right)\leq O\left(\left(F_{\vert H_{ii}\vert ^2}\left(  \frac{\alpha_n^2\sigma^2_N}{\bar p }\right)\right)^2\right)
\end{equation}
and by combining it with \eqref{20_86}, \eqref{20_87} and  \eqref{babababa} we obtain
the desired result. \hfill $\Box$

\subsection{Proof of Proposition \ref{theorem that coopetation dominates}}
\label{appendix_0}
Player i's conditional expected payoff is
\beqna
\pi_i( S_i, S_j,x_i,y_i)=\max\left\{{ a_ju_i(1,1,x_i,y_i)+ (1- a_j)u_i(1,1/2,x_i,y_i)}\right. \\ ,\left.{ a_ju_i(1/2,1,x_i,y_i)+ (1- a_j)u_i(1/2,1/2,x_i,y_i)}\right\}
\eeqna
where $a_j=P(S_j=1)$. Thus, it is sufficient to show that
\beqna
 a_ju_i(1/2,1,x_i,y_i)+ (1- a_j)u_i(1/2,1/2,x_i,y_i) > u_i(1/2,1/2,x_i,y_i),\;\forall x_i,y_i\in {\cal X}_i\times{\cal Y}_i
\eeqna
This is equivalent to
\beqna
\lefteqn{\frac{y_i^3 x_i^2}{2}+2 y_i^3 x_i+\frac{y_i^2
   x_i^3}{2}+4 y_i^2 x_i^2+8 y_i^2
   x_i+\frac{x_i^4y_i}{8}}\nonumber\\&&+2  x_i^3y_i+9
   x_i^2y_i+12 y_i x_i+\frac{x_i^4}{4}+2 x_i^3+6
   x_i^2+6 x_i> 0
\eeqna
which is always true.
\hfill $\Box$
\subsection{Proof of Theorem \ref{approximate nash points representation theorem}}
\label{approximate nash points representation theorem - proof}

We begin with the following definition:
\begin{definition}
\label{approximate NE definition}
An approximate NE point is the strategy profile $ ( \tilde{S}_1(x_1,y_1,\hat a_2),\tilde{S}_2(x_2,y_2,\hat a_1))$
where $\hat a_1$ and $\hat a_2$ are a solution to  equations  (\ref{a_1 curve}) and (\ref{a_2 curve}).
\end{definition}

It remains to show that if there exists an approximate NE point, then there exists a $\epsilon$-near NE point  given by (\ref{approximate nash points representation theorem - equation1}) and (\ref{approximate nash points representation theorem - equation2}). Let
\beqna
\tilde a_j=P\left(\check{S}_j\left(X_i,Y_i,\hat a_i \right)=1 \right)\\
\tilde a_i=P\left(\check{S}_i\left(X_i,Y_i,\tilde a_j \right)=1 \right)
\eeqna
 In words, $\tilde a_j$ is the probability that player $j$    chooses  FDM  if he is not deviating  from the $\epsilon$-near NE point and   $\tilde a_i$ is the probability that player $i$  chooses FDM if he ``cheats" and uses his best response to player $j$'s true probability for choosing FDM $\tilde a_i$ rather than the probability $\hat a_j$.

To show that $\left(\check S_i\left(x_i,y_i,\hat a_j\right),\check S_j\left(x_j,y_j,\hat a_i\right)\right)$ satisfies (\ref{defination near optimal nash points equation 1}), one needs to show that for every $x_i,y_i$ $\in$ ${\cal X}_i\times{\cal Y}_i$ and for sufficiently large $\bar p$
\beqna
\Delta \pi_i(x_i,y_i)=\vert \pi_i\left(\check S_i\left(x_i,y_i,\tilde a_j\right),\check S_j\left(x_j,y_j,\hat a_i\right)\right)-\pi_i\left(\check S_i\left(x_i,y_i,\hat a_j\right),\check S_j\left(x_j,y_j,\hat a_i\right)\right)\vert<\epsilon
\eeqna
Note that $\Delta \pi_i(x_i,y_i)\neq 0$ if and only if $(x_i,y_i)\in\check{D}^{\hat a_j}\Delta \check D_i^{\tilde a_j}$ (since player $j$'s true probability for choosing FDM is identical in both cases and is equal to $\tilde a_j$), thus
\beqna
\Delta \pi_i(x_i,y_i,\hat a_j,\tilde a_j)=\vert e(x_i,y_i,{\tilde a_i})\vert I_{\check{D}^{\hat a_j}\Delta \check D_i^{\tilde a_j}}(x_i,y_i)
\eeqna
where $I_A(x,y)$ denotes the indicator function, i.e. it is equal to $1$ if $(x,y)\in A$ and zero otherwise. Since $ (x_i,y_i)\in\check{D}^{\hat a_j}\Delta \check D_i^{\tilde a_j}$ is equivalent to $e(x_i,y_i,\hat a_j)>0$ and $e(x_i,y_i,\tilde a_j)\leq 0$ or vice versa, and because $e(x,y,a) $   is a continuous function of $a$,  for every \(\hat a_j,\tilde a_j\) there exists  some $a^*$ in the interval between $\hat a_j$ and $\tilde a_j$ such that $e(x_i,y_i, a^*)=0$. By Lemma \ref{equivalence_condition}, we know that $\tilde a_j \xrightarrow[\bar{p}\to \infty ]{} \hat a_j$,  thus $e(x_i,y_i,\tilde a_j ) \xrightarrow[\bar{p}\to \infty ]{}0$, furthermore because $e(x,y,a)$ is bounded for every \(x,y\) and is  a linear function of \textit{a} it follows that \beqna
\vert e(x_i,y_i,{\tilde a_i})\vert I_{\check{D}^{\hat a_j}\Delta \check D_i^{\tilde a_j}}(x_i,y_i)=O\left( \tilde a_i-\hat a_j \right)
\label{17_48}\eeqna
\hfill $\Box$

\subsection{Proof of Theorem \ref{theorem 3}}
\label{proof_theorem_exstans}
Denote $w_i(a_j)=1-F_{Z_i}(q(a_j))$ for $i\neq j$.  Thus
\beq
\label{27}
w'_i(a_j)=-f_{Z_i}(q(a_j))q'(a_j)
\eeq
Before analyzing \eqref{27} recall that $\lim q(a)_{a \rightarrow 0}=\infty$, furthermore, it can be verified that \beq
\label{equation of appendix 2}
\lim_{a\rightarrow 0}\frac{q'(a)}{q^2(a)\log(q(a))}= M
\eeq
(this follows immediately from \eqref{19}.
Thus, if (\ref{25}) is satisfied
\beq \lim_{a_j\rightarrow 0} w'(a_j)=\infty
\eeq
Consider the curves (\ref{a_1 curve}) and (\ref{a_2 curve}) in a two-dimensional cartesian   system where $a_1$ and $a_2$ are given by the horizontal and the vertical coordinates respectively.  Both curves are continuous and differentiable. Furthermore, the point $(0,0)$ is a common point of the two curves and the points $(1-F_{Z_1}(0.5),1)$, $(1,1-F_{Z_2}(0.5))$ lie on curves (\ref{a_1 curve}) and (\ref{a_2 curve}) respectively. Since the slop of curve (\ref{a_1 curve}) tends to zero as $a_1 \rightarrow 0$ and the slope of curve (\ref{a_2 curve}) tends to infinity as $a_1\rightarrow 0$, the two curves must intersect at least once.
\hfill $\Box$
\bibliographystyle{ieeetr}

\end{document}